\documentclass[pre,aps,floatfix,twocolumn]{revtex4-1}
\usepackage{amsmath,amssymb,multirow,epsfig,bm,pifont}
\usepackage{mathtools, float}
\usepackage[toc,page]{appendix}
\usepackage{comment}

\newcommand{\beq}{\begin{equation}}
\newcommand{\eeq}{\end{equation}}
\newcommand{\bea}{\begin{eqnarray}}
\newcommand{\eea}{\end{eqnarray}}

\newcommand{\Var}{\text{Var}}

\usepackage[plainpages=false,pdfpagelabels,colorlinks=true,linkcolor=red,urlcolor=blue,citecolor=blue,pdftitle={},pdfauthor={},pdfdisplaydoctitle=true,pdfduplex=DuplexFlipLongEdge]{hyperref}

\begin{document} 

\title{Eigenstate thermalization for observables that break Hamiltonian symmetries\\ and its counterpart in interacting integrable systems}
\author{Tyler LeBlond}
\affiliation{Department of Physics, The Pennsylvania State University, University Park, Pennsylvania 16802, USA}
\author{Marcos Rigol}
\affiliation{Department of Physics, The Pennsylvania State University, University Park, Pennsylvania 16802, USA}

\begin{abstract}
We study the off-diagonal matrix elements of observables that break the translational symmetry of a spin-chain Hamiltonian, and as such connect energy eigenstates from different total quasimomentum sectors. We consider quantum-chaotic and interacting integrable points of the Hamiltonian, and focus on average energies at the center of the spectrum. In the quantum-chaotic model, we find that there is eigenstate thermalization; specifically, the matrix elements are Gaussian distributed with a variance that is a smooth function of $\omega=E_{\alpha}-E_{\beta}$ ({$E_{\alpha}$} are the eigenenergies) and scales as $1/D$ ($D$ is the Hilbert space dimension). In the interacting integrable model, we find that the matrix elements exhibit a skewed log-normal-like distribution and have a variance that is also a smooth function of $\omega$ that scales as $1/D$. We study in detail the low-frequency behavior of the variance of the matrix elements to unveil the regimes in which it exhibits diffusive or ballistic scaling. We show that in the quantum-chaotic model the behavior of the variance is qualitatively similar for matrix elements that connect eigenstates from the same versus different quasimomentum sectors. We also show that this is not the case in the interacting integrable model for observables whose translationally invariant counterpart does not break integrability if added as a perturbation to the Hamiltonian.
\end{abstract}

\maketitle

\section{Introduction}

The emergence of thermalization under unitary dynamics in generic isolated quantum systems has been intensively explored over the past decade~\cite{ETH_review, mori_ikeda_18, eisert_friesdorf_review_15, polkovnikov_sengupta_review_11}. On the experimental side, where high levels of control and isolation in ultracold atomic gases have recently enabled the study of quantum dynamics over long time scales~\cite{Langen2015, Bloch2012, Maciej2007}, both thermalization and the lack thereof have been observed in chaotic~\cite{Trotzky2012, kaufman_tai_16, clos_porras_16, tang2018} and (near-)integrable~\cite{Kinoshita2006, gring_kuhnert_12, langen_erne_15, tang2018} quantum systems, respectively. Thermalization in quantum chaotic systems is generally understood in the context of the eigenstate thermalization hypothesis (ETH)~\cite{ETH_review, Deutsch1991, Srednicki1994, ETHRigol}. On the integrable side, thermalization is precluded by an extensive set of local conserved quantities, though equilibration in these systems has also been the subject of much interest~\cite{GGE_review, GGE_Rigol_2007, wouters_denardis_14, pozsgay_mestyan14, ilievski_denardis_15}.  

As the outcomes of quantum dynamics are ultimately determined by properties of matrix elements, the content of the ETH is usually expressed through a matrix-element ansatz for few-body operators (observables) in the eigenstates of chaotic Hamiltonians~\cite{Srednicki_1999, ETH_review}:
\beq \label{eq:ETH} 
O_{\alpha\beta}=O(\bar{E})\delta_{\alpha\beta}+e^{-S(\bar{E})/2}f_O(\bar{E},\omega)R_{\alpha\beta},
\eeq
where the average energy $\bar{E}\equiv (E_{\alpha}+E_{\beta})/2$, the frequency $\omega=E_{\alpha}-E_{\beta}$, and $S(\bar E)$ is the thermodynamic entropy at energy $\bar E$. The functions $O(\bar{E})$ and $f_O(\bar{E},\omega)$ are smooth, and $R_{\alpha\beta}$ is a Gaussian distributed variable with zero mean and unit variance (variance 2) for $\alpha \neq \beta$ ($\alpha=\beta$) in Hamiltonians that exhibit time-reversal symmetry, namely, in Hamiltonians that can be represented by real matrices. Hence, the ETH states that the diagonal matrix elements of observables in the energy eigenbasis are smooth functions of the energy. This is what makes thermalization (the agreement between long-time results and statistical mechanics predictions) possible. The ETH also states that the off-diagonal matrix elements are exponentially small, and this ensures equilibration (the time fluctuations of observables about the time average are small) at long times~\cite{ETH_review}. The smooth function $|f_O(\bar{E},\omega)|^2$ is central to fluctuation-dissipation relations~\cite{ETH_review}, and can be probed experimentally by measuring heating rates in periodically driven systems~\cite{heatingrates}.

In quantum integrable systems, the presence of extensive sets of local conserved quantities is manifest in the properties of the matrix elements of observables. It is known that the diagonal matrix elements have both a support that does not vanish in the thermodynamic limit and average fluctuations that decay as a power law in system size~\cite{Breakdown, rigol_09b, PhysRevLett.105.250401, Santos2010Localization, GGE_Rigol_2011, Steinigeweg2013Eigenstate, Ikeda, Beug1, Alba, GGE_review, 2DTFIM, yoshizawa2018numerical, PRE2019, mierzejewski_vidmar_19}, and thus defy Eq.~(\ref{eq:ETH}). In an interacting integrable system (the spin-1/2 XXZ chain), the off-diagonal matrix elements were recently found to be nearly log-normally distributed~\cite{PRE2019}. In addition, it was found that the variance is a smooth function of $\omega$ (for $\bar{E}$ at the center of the spectrum) that scales as prescribed by the ETH (as a result, it can also be probed experimentally by measuring heating rates in periodically driven systems~\cite{heatingrates}). The scaling of other moments, of course, is not determined by the scaling of the variance, which means that there is no equivalent of the off-diagonal part of Eq.~\eqref{eq:ETH} in integrable models.

Using that in interacting integrable systems one can define a smooth scaled variance $V_O(\bar{E},\omega) \equiv e^{S(\bar{E})} \Var(O_{\alpha\beta})$~\cite{PRE2019}, recent works have unveiled properties of that function at low values of $\omega$ (for $\bar{E}$ at the center of the spectrum of spin-1/2 lattice Hamiltonians, $\bar{E}\approx0$)~\cite{brenes2020eigenstate, p2020adiabatic, brenes2020ballistic, Chandran20}. Via the computation of the adiabatic gauge potential (AGP) norm, in Ref.~\cite{p2020adiabatic} it was shown that at exponentially small (in system size) frequencies $V_O(0,\omega)$ vanishes for observables that do not break integrability if added as perturbations to the Hamiltonian, while it scales as in quantum chaotic models for observables that do. Such behaviors were observed in Ref.~\cite{brenes2020ballistic} at frequencies that are polynomially small in the system size. There it was also shown that observables for which $V_O(0,\omega\rightarrow0)$ scales as in quantum chaotic models do not exhibit eigenstate thermalization at integrability.

By now, several studies have explored properties of the off-diagonal matrix elements of observables in quantum chaotic~\cite{ETH_review, ETHRigol, rigol_09b, Santos2010Localization, Khatami, Steinigeweg2013Eigenstate, Beug2, 2DTFIM-2, khaymovich_haque_19, PRE2019, Jansen2019Eigenstate, Dymarsky2019a, Dymarsky2019b, brenes2020eigenstate, p2020adiabatic, brenes2020ballistic, Chandran20, speckLea, Richter2020} and integrable~\cite{rigol_09b, Santos2010Localization, Khatami, Beug2, PRE2019, Dymarsky2019a, Dymarsky2019b, brenes2020eigenstate, p2020adiabatic, brenes2020ballistic, Chandran20, speckLea, Richter2020} models. In this work we aim to contribute to that existing body of literature by studying the off-diagonal matrix elements (in the energy eigenbasis) of observables that break symmetries of the Hamiltonian. Specifically, we study the off-diagonal matrix elements of observables that break translational symmetry in the eigenstates of translationally invariant Hamiltonians. This means that the off-diagonal matrix elements are nonvanishing between eigenstates from different total quasimomentum sectors. We are not aware of previous studies of the structure of such matrix elements.

We compute these matrix elements in the eigenstates of both a quantum-chaotic model and an interacting integrable model, for average energies at the center of the spectrum. In the quantum-chaotic model, we find that the off-diagonal matrix elements exhibit all of the properties prescribed by the ETH. We also find that finite-size effects are larger in matrix elements that connect eigenstates from different total quasimomentum sectors (the overwhelming majority of the matrix elements) than in matrix elements that connect eigenstates from the same quasimomentum sector. Since, for eigenstates from the same quasimomentum sector, the matrix elements of operators that break translational symmetry are identical to those of the corresponding translationally invariant operator, another way to phrase the latter finding is that nontranslationally invariant observables exhibit larger finite-size effects than their translationally invariant counterparts. In the interacting integrable model, we find that the distribution of the matrix elements of the nontranslationally invariant observables is skewed log-normal-like with zero mean and a variance that scales as $1/D$ ($D$ is the Hilbert space dimension), as found in Ref.~\cite{PRE2019} for translationally invariant observables.

Another major goal of this work is to understand the low-frequency behavior of the scaled variances. For quantum-chaotic systems, for which the ETH~\eqref{eq:ETH} is expected to be valid, we refer to the scaled variances as $|f_O(\bar{E},\omega)|^2$. For integrable systems, for which there is no well defined $f_O(\bar{E},\omega)$ function (the scaling of the moments of the distribution of $O_{\alpha\beta}$ is not determined by the scaling of the variance, as mentioned before), we refer to the scaled variances as $V_O(\bar{E},\omega)$. We focus on $\bar{E}$ at the center of the spectrum ($\bar{E}\approx0$), which is where the overwhelming majority of matrix elements is located in our local Hamiltonians. In the quantum-chaotic model, we find $|f_O(0,\omega)|^2$ to be consistent with random matrix theory, namely, to exhibit a plateau as $\omega\rightarrow0$ (with a diffusive scaling)~\cite{ETH_review}. In the interacting integrable model, we find the behavior and scaling of $V_O(0,\omega)$ to be rich and observable dependent. For matrix elements that connect energy eigenstates from within the same total quasimomentum sector, we find two possible behaviors as $\omega\rightarrow0$. Either $V_O(0,\omega)$ goes to a nonzero value proportional to $L$ (as in quantum-chaotic models), or it vanishes. For matrix elements that connect energy eigenstates from different quasimomentum sectors, we find that $V_O(0,\omega)$ always goes to a nonzero value proportional to $L$. Hence, there are observables for which the $\omega\rightarrow0$ behavior of $V_O(0,\omega)$ is qualitatively different between matrix elements that connect energy eigenstates from the same quasimomentum sector and those that connect eigenstates from different quasimomentum sectors. In Sec.~\ref{integ}, we discuss the connection between these findings and the results in Refs.~\cite{brenes2020eigenstate, p2020adiabatic, brenes2020ballistic, Chandran20}.

The presentation is organized as follows: In Sec.~\ref{model}, we introduce the spin-1/2 chains and the specific observables studied, and discuss details of our numerical calculations. In Sec.~\ref{chaos}, we report our results for the off-diagonal matrix elements of observables in the quantum-chaotic model, which include a characterization of their distributions and the study of their variances. In Sec.~\ref{integ}, we carry out a parallel analysis for the interacting integrable model. In Sec.~\ref{conc}, we summarize our results. 

\section{Model}\label{model}

We study the same spin-1/2 chains as in Ref.~\cite{PRE2019}, namely, the XXZ chain with the addition of next-nearest neighbor interactions and periodic boundary conditions. The Hamiltonian reads
\begin{eqnarray}\label{eq:xxz} 
\hat{H} &=& \sum_{i=1}^L \left[ \frac{1}{2} \left( \hat{S}_i^+\hat{S}_{i+1}^- + \text{H.c.} \right) + \Delta \hat{S}_i^z\hat{S}_{i+1}^z \right]\nonumber \\
&&+ \lambda \sum_{i=1}^L \left[ \frac{1}{2} \left( \hat{S}_i^+\hat{S}_{i+2}^- + \text{H.c.} \right) + \frac{1}{2} \hat{S}_i^z\hat{S}_{i+2}^z \right] ,
\end{eqnarray}
where $\hat{S}_i^{\nu}$ are spin-1/2 operators in the $\nu\in\{x,y,z\}$ directions on site $i$ (represented by Pauli matrices), $\hat{S}_i^{\pm} = \hat{S}_i^x\pm i\hat{S}_i^y$ are the corresponding ladder operators, and $L$ is the number of lattice sites. $\Delta$ is the so-called anisotropy parameter in the XXZ chain, and $\lambda\neq0$ breaks the integrability of the XXZ chain~\cite{PhysRevE.81.036206}. In Sec.~\ref{chaos}, we set $\lambda=1$ to study matrix elements of Hamiltonian~\eqref{eq:xxz} in the quantum-chaotic regime, while in Sec.~\ref{integ} we set $\lambda=0$ to study matrix elements at integrability. We mostly compare results for $\Delta=0.55$ (easy-plane regime of the XXZ chain) and $\Delta=1.1$ (easy-axis regime of the XXZ chain).

To study the matrix elements of observables in the energy eigenstates of Hamiltonian~\eqref{eq:xxz}, it is important to resolve all of its symmetries~\cite{ETH_review, Santos2010Localization}. First, we note that the Hamiltonian commutes with $\hat M^z = \sum_i\hat{S}_i^z$, which is the total magnetization in the $z$-direction. We focus on the zero magnetization sector of chains with an even number of lattice sites. This sector has an additional spin inversion ($Z_2$) symmetry; we focus on the even-$Z_2$ sector. Next, translational symmetry allows us to block-diagonalize the Hamiltonian in different total quasimomentum $k$ sectors. Lastly, within the $k=0$ and $\pi$ sectors, we resolve the space reflection ($P$) symmetry.

We study the matrix elements of three local operators that break the translation symmetry of Hamiltonian~\eqref{eq:xxz}: the nearest-neighbor $z$-interaction
\beq \label{eq:obs_Unn}
\hat{U}_\text{n}=\hat{S}^z_{1}\hat{S}^z_{2},
\eeq
the next-nearest-neighbor $z$-interaction
\beq \label{eq:obs_Unnn}
\hat{U}_\text{nn}=\hat{S}^z_{1}\hat{S}^z_{3},
\eeq
and the next-nearest-neighbor flip-flop operator
\beq \label{eq:obs_Knnn}
\hat{K}_\text{nn}=\hat{S}^+_1\hat{S}^-_{3}+\hat{S}^+_{3}\hat{S}^-_1.
\eeq
These local operators connect all total quasimomentum sectors of the Hamiltonian. Since the Hamiltonian is translationally invariant, the sites used to define $\hat{U}_\text{n}$, $\hat{U}_\text{nn}$, and $\hat{K}_\text{nn}$ do not influence the results.

The first important consequence of the translational symmetry of the Hamiltonian is that the diagonal matrix elements of $\hat{U}_\text{n}$, $\hat{U}_\text{nn}$, and $\hat{K}_\text{nn}$ (referred to in what follows as ``symmetry-breaking'' operators) are identical to the diagonal matrix elements of the corresponding translationally invariant operators (referred to in what follows as ``symmetry-preserving'' operators)
\bea
\hat{U}^T_\text{n}&=&\frac{1}{L}\sum_{i=1}^L\hat{S}^z_{i}\hat{S}^z_{i+1},\label{eq:UNNT}\\
\hat{U}^T_\text{nn}&=&\frac{1}{L}\sum_{i=1}^L\hat{S}^z_{i}\hat{S}^z_{i+2},\\
\hat{K}^T_\text{nn}&=&\frac{1}{L}\sum_{i=1}^L\left(\hat{S}^+_i\hat{S}^-_{i+2}+\hat{S}^+_{i+2}\hat{S}^-_i\right).\label{eq:KNNNT}
\eea
In addition, within a given total quasimomentum sector, the off-diagonal matrix elements of $\hat{U}_\text{n}$, $\hat{U}_\text{nn}$, and $\hat{K}_\text{nn}$ are identical to those of $\hat{U}^T_\text{n}$, $\hat{U}^T_\text{nn}$, and $\hat{K}^T_\text{nn}$, respectively.

The diagonal and the off-diagonal matrix elements of symmetry-preserving operators were studied in detail (within the $k=0$ sector) in Ref.~\cite{PRE2019}. In this work our focus will be on off-diagonal matrix elements. In our discussions, by way of comparing the set of all matrix elements with the set of matrix elements that connect energy eigenstates from the same quasimomentum sector, we contrast the behaviors of matrix elements of the symmetry-breaking operators with those of their symmetry-preserving counterparts, respectively.

The off-diagonal matrix elements ($O_{\alpha\beta}$) of symmetry-breaking operators in the energy eigenstates are obtained using full exact diagonalization within the even-$Z_2$ sector of the $M^z=0$ sector (with dimension $D^e_{Z_2}$) that, in turn, is split in $L$ total quasimomentum $k$ sectors. Whenever $k_\alpha$ or $k_\beta$ are neither 0 nor $\pi$, one generally has $O_{\alpha\beta}\neq0$. For the off-diagonal matrix elements within the $k=0,\,\pi$ sectors and between them, for which space reflection symmetry is resolved, we remove from our analyses the blocks of matrix elements that are zero~\footnote{Within the $k=0$ ($k=\pi$) sector, the off-diagonal matrix elements vanish between blocks with different parity (under space reflection). Between the $k=0$ and $\pi$ sectors, the off-diagonal matrix elements of $\hat{U}_\text{n}$ vanish between blocks with different parity, while they vanish for $\hat{U}_\text{nn}$ and $\hat{K}_\text{nn}$ between blocks with the same parity.}. We note that, when reporting the $k_\alpha=k_\beta$ results for $O_{\alpha\beta}$, we exclude the $k=0$ and $\pi$ sectors (two out of $L$ sectors) due to the extra symmetry present in those sectors.

The dimension $D$ of the Hilbert space used in our normalization for each observable is the square root of the total number of matrix elements that do not vanish for symmetry reasons. Since the number of blocks with vanishing off-diagonal matrix elements is only $O(1)$, $D\simeq D^e_{Z_2}$. We carry out calculations for chains with up to $L=22$, including all quasimomentum sectors. For matrix elements that connect eigenstates from the same quasimomentum sectors, we carry out calculations up to $L=24$. For our low-frequency analyses, we also report results for the $k=0$, even-$P$, even-$Z_2$ sector up to $L=26$~\cite{PRE2019}. 

\section{Quantum-Chaotic Chain} \label{chaos}

In this section, we study the off-diagonal matrix elements of our observables of interest in the eigenstates of Hamiltonian~\eqref{eq:xxz} with $\Delta=0.55,\,1.1$, and $\lambda=1$. We focus on the regime $\bar{E}=(E_{\alpha}+E_{\beta})/2\approx0$, namely, on average energies at the center of the spectrum (the so-called infinite-temperature regime).

\subsection{Distributions}

Here we characterize the distribution of $|O_{\alpha\beta}|$. We take the absolute value because $O_{\alpha\beta}$ is complex whenever $k_\alpha$ or $k_\beta$ are neither 0 nor $\pi$. In addition to considering $\bar{E} \approx 0$, we first focus on the regime in which $\omega = |E_{\alpha}-E_{\beta}| \approx 0$. In the context of the ETH ansatz, $|f_O(\bar{E},\omega)|$ exhibits a plateau in this regime~\cite{ETH_review}, and the distribution of $|O_{\alpha\beta}|$ is expected to be the same as in random matrix theory.

\begin{figure*}[!t]
\centering
\includegraphics[width=0.9\textwidth]{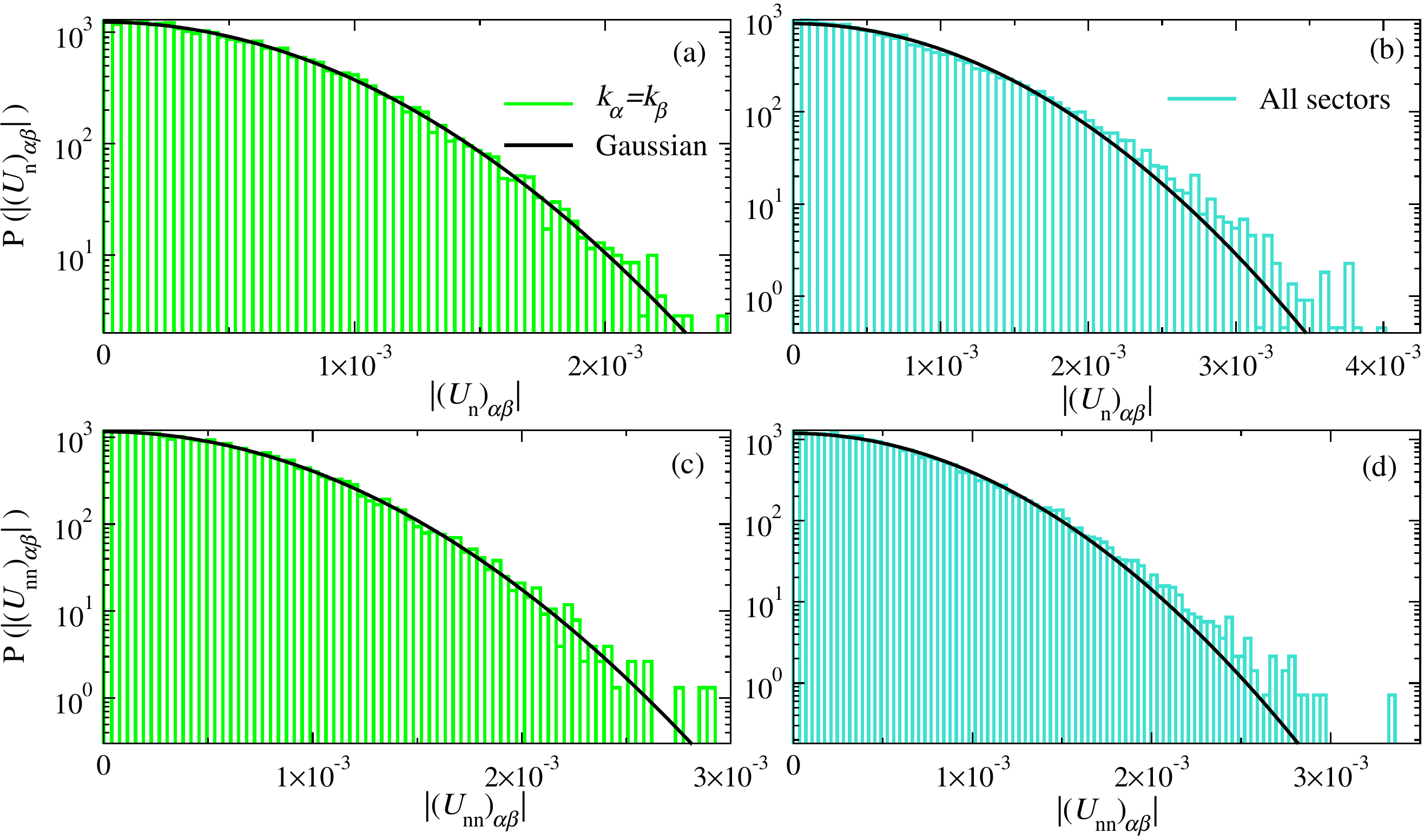}
\vspace{-0.1cm}
\caption{\label{fig:1} Probability distributions $P(|O_{\alpha\beta}|)$ for observables $\hat{U}_\text{n}$ [(a), (b)] and $\hat{U}_\text{nn}$ [(c), (d)] for Hamiltonian~\eqref{eq:xxz} with $\Delta=0.55$ (similar results were obtained for $\Delta=1.1$) and $\lambda=1$ (quantum-chaotic regime). We consider pairs of energy eigenstates for which $|\bar{E}|/L \leq 0.025$, and choose the 40,000 matrix elements with the lowest $\omega$ (this results in $\omega \leq 0.001$). We show results for matrix elements with $k_{\alpha}=k_{\beta}$ (excluding the $k=0$ and $\pi$ sectors, we do the same in all plots that follow) [(a), (c)] and matrix elements that mix all quasimomentum sectors [(b), (d)], in the $L=22$ chain. The continuous lines are half-normal distributions with the same variance as the distributions $P(|O_{\alpha\beta}|)$.} \end{figure*} 

Figure~\ref{fig:1} shows the probability distributions of $|(U_\text{n})_{\alpha\beta}|$ [(a), (b)] and $|(U_\text{nn})_{\alpha\beta}|$ [(c), (d)] for Hamiltonian~\eqref{eq:xxz} with $\Delta=0.55$, in a chain with $L=22$ (qualitatively similar results were obtained, not shown, for $|(K_\text{nn})_{\alpha\beta}|$). Figures~\ref{fig:1}(a) and~\ref{fig:1}(c) show the distributions for pairs of energy eigenstates with $k_{\alpha}=k_{\beta}$, and Figs.~\ref{fig:1}(b) and~\ref{fig:1}(d) show the distributions for pairs that connect all quasimomentum sectors. In all panels in Fig.~\ref{fig:1} we also show half-normal distributions, for which the variances are the same as those of the numerical results, as continuous black lines.

Overall, the results in Fig.~\ref{fig:1} show that $|(U_\text{n})_{\alpha\beta}|$ and $|(U_\text{nn})_{\alpha\beta}|$ are normally distributed regardless of whether one looks at eigenstate pairs for which $k_\alpha=k_\beta$ (i.e., at symmetry-preserving operators) or at all eigenstate pairs (i.e., at symmetry-breaking operators). A comparison between the results in the left columns ($k_\alpha=k_\beta$) and the right columns (all eigenstate pairs) of Fig.~\ref{fig:1} suggests that the variances of the distributions are generally different between the symmetry-preserving and the symmetry-breaking versions of any given observable, and that the magnitude of the difference depends on the observable. We continue to explore those observations in the next subsections.

Next, we probe the Gaussianity of the distributions of matrix elements for $\omega>0$. For that purpose we calculate the ratio~\cite{PRE2019}
\beq \label{eq:ratio} 
\Gamma_O(\omega)=\overline{|O_{\alpha\beta}|^2}/\overline{|O_{\alpha\beta}|}^2.
\eeq 
In Eq.~\eqref{eq:ratio}, $\overline{(\dots)}$ denotes a coarse-grained average (over small $\delta\omega$ windows) for pairs of energy eigenstates that satisfy $|\bar{E}|/L\le0.025$. If $O_{\alpha\beta}$ has a Gaussian distribution with zero mean, then $\Gamma_O(\omega)=\pi/2$. $\Gamma_O(\omega)$ has been computed recently for various models and observables~\cite{PRE2019, brenes2020ballistic, brenes2020eigenstate, speckLea}, as the normality of the distribution of off-diagonal matrix elements of observables has been used to probe eigenstate thermalization.

\begin{figure}[!b]
\centering \includegraphics[width=\columnwidth]{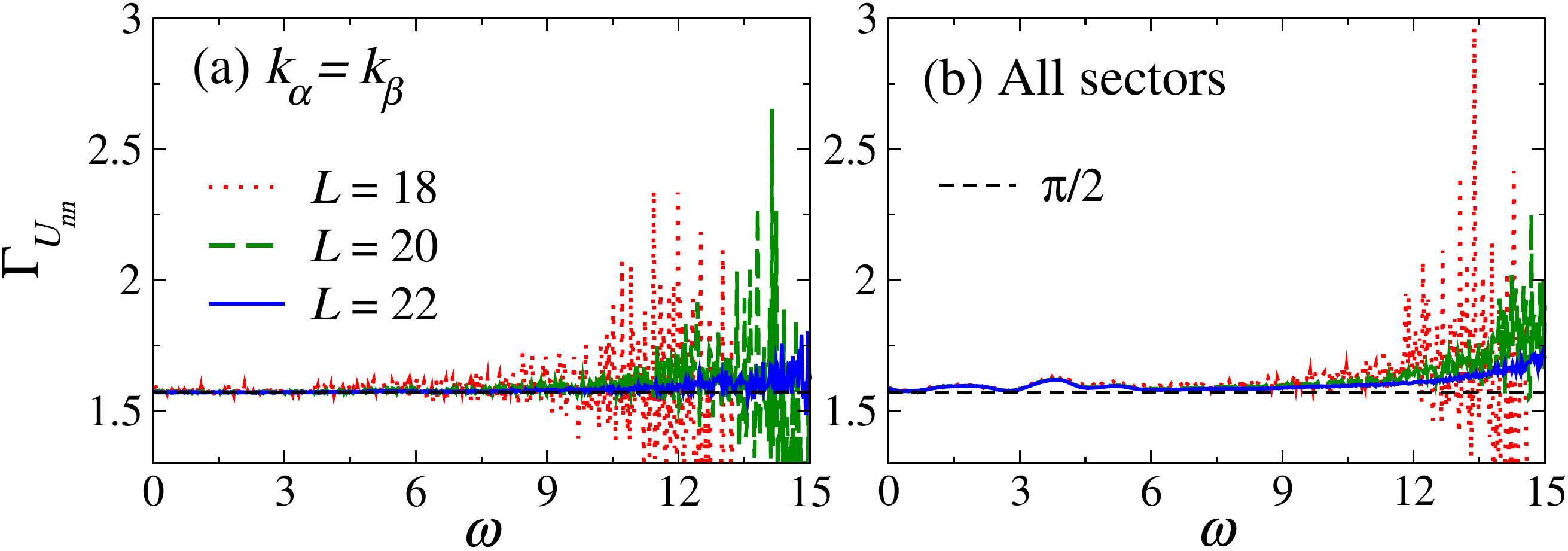}
\vspace{-0.5cm}
\caption{\label{fig:2} $\Gamma_{U_\text{nn}}$ [see Eq.~\eqref{eq:ratio}] at a nonintegrable point ($\lambda=1$) of Hamiltonian~\eqref{eq:xxz} with $\Delta=0.55$ for different chain sizes (similar results were obtained for $\Delta=1.1$). We show results for pairs of energy eigenstates with $k_{\alpha}=k_{\beta}$ (a) and pairs that mix all quasimomentum sectors (b). All pairs of eigenstates satisfy $|\bar{E}|/L\leq 0.025$. The averages $\overline{|(U_\text{nn})_{\alpha\beta}|}$ and $\overline{|(U_\text{nn})_{\alpha\beta}|^2}$ were coarse-grained in windows of width $\delta \omega = 0.025$.}
\end{figure}

In Fig.~\ref{fig:2}(a), we show $\Gamma_{U_\text{nn}}(\omega)$ for matrix elements that connect energy eigenstates with the same quasimomentum ($k_{\alpha}=k_{\beta}$), and in Fig.~\ref{fig:2}(b) we show $\Gamma_{U_\text{nn}}(\omega)$ for matrix elements that connect all sectors. The results in Fig.~\ref{fig:2}(a) appear to have converged to $\Gamma_{U_\text{nn}}(\omega)=\pi/2$, with deviations at large values of $\omega$ occurring because of finite-size effects (the curves move toward $\pi/2$ with increasing $L$). Figure~\ref{fig:2}(b) contains deviations from Gaussianity (small bumps) for $\omega<6$, but overall exhibits the same behavior as Fig.~\ref{fig:2}(a). Both other observables we studied ($\hat{U}_\text{n}$ and $\hat{K}_\text{nn}$) exhibited qualitatively similar behaviors for both sets of matrix elements, indicating that, for the chain sizes accessible to us: (i) the distributions of matrix elements appear to be Gaussian at all frequencies and (ii) finite-size effects (in the form of deviations from Gaussianity at intermediate values of $\omega$) are stronger for symmetry-breaking observables than for symmetry-preserving ones.

In Ref.~\cite{PRE2019}, for translationally invariant observables within the $k=0$ quasimomentum sector, a small nearly $L$-independent deviation in $\Gamma_O(\omega)$ from $\pi/2$ was observed for $5\lesssim\omega\lesssim 8$ (for the chain sizes available). That deviation was argued to be consistent with strong finite-size effects. In Fig.~\ref{fig:2}(a), which includes results from all pairs of energy eigenstates with $k_{\alpha}=k_{\beta}$, one can see that $\Gamma_{U_\text{nn}}(\omega)$ approaches $\pi/2$ with increasing $L$ in that frequency regime. This further strengthens the case that the deviations from Gaussianity seen in Ref.~\cite{PRE2019} for translationally invariant observables are the result of finite-size effects. In Fig.~\ref{fig:2}(b), and for $\hat{U}_\text{n}$ and $\hat{K}_\text{nn}$ (not shown), we see similar small nearly $L$-independent deviations from $\pi/2$. No such deviations have been observed in recent full exact diagonalization calculations in systems with broken translational symmetry~\cite{brenes2020ballistic, brenes2020eigenstate}, so we attribute them here to strong finite-size effects for symmetry-breaking observables in our translationally invariant energy eigenstates. To further test this, we performed calculations for larger [but still $O(1)$] values of $\lambda$ and found that the deviations from $\pi/2$ decrease deeper in the quantum chaotic regime.

\subsection{Variances}\label{sec:variancesQC}

Next we study the behavior of the off-diagonal matrix elements and their variances as functions of the frequency $\omega$, as well as the scaling of the variances with system size. Since the average $\overline{O_{\alpha\beta}}=0$, the variances are given by the averages $\overline{|O_{\alpha\beta}|^2}$, namely, $\Var(O_{\alpha\beta})=\overline{|O_{\alpha\beta}|^2}$. 

\begin{figure}[!b]
\centering \includegraphics[width=\columnwidth]{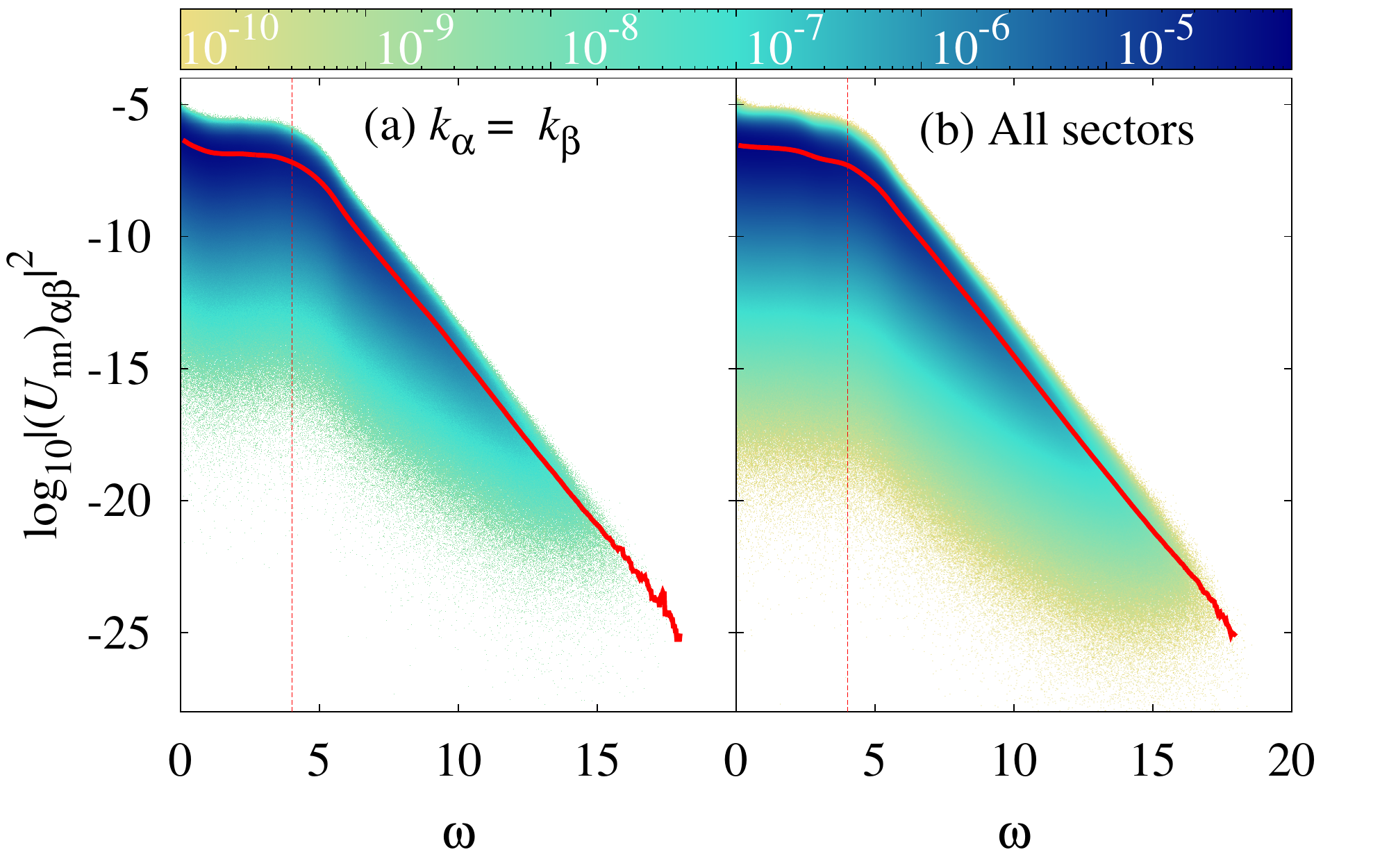}
\vspace{-0.4cm}
\caption{\label{fig:3} Normalized 2D histograms of $\log_{10}|(U_\text{nn})_{\alpha\beta}|^2$ vs $\omega$ at a nonintegrable ($\lambda=1$) point of Hamiltonian~\eqref{eq:xxz} with $\Delta=0.55$ for $L=22$ (qualitatively similar results were obtained for $\Delta=1.1$). We consider pairs of energy eigenstates with $k_{\alpha}=k_{\beta}$ (a) and pairs that mix all quasimomentum sectors (b). All pairs of energy eigenstates satisfy $|\bar{E}|/L\le0.025$. The (red) solid lines are running averages $\log_{10}\overline{|(U_\text{nn})_{\alpha\beta}|^2}$ calculated in windows of width $\delta\omega=0.175$ centered at points separated by $\Delta\omega=0.025$. The vertical dashed lines show the values of $\omega$ up to which results for $|O_{\alpha\beta}|^2$ are included in the scaling analysis of Fig.~\ref{fig:4}.}
\end{figure}

In Fig.~\ref{fig:3}, we visualize the distribution of $\log_{10}{|(U_\text{nn})_{\alpha\beta}|^2}$ as a function of $\omega$ using normalized 2D histograms for matrix elements between pairs of energy eigenstates with $k_{\alpha}=k_{\beta}$ [Fig.~\ref{fig:3}(a)] and between pairs that connect all quasimomentum sectors [Fig.~\ref{fig:3}(b)]. In both panels, we have included matrix elements for pairs of energy eigenstates for which $\bar{E}/L \leq 0.025$, and used $\Delta=0.55$ for chains with $L=22$. The results are qualitatively similar in Fig.~\ref{fig:3}(a) and~\ref{fig:3}(b), and they are qualitatively similar to the results for translationally invariant operators in the $k=0$ sector reported in Ref.~\cite{PRE2019}. This reveals that the matrix elements of symmetry-breaking operators are not qualitatively affected by the block diagonal structure of the Hamiltonian matrix.

In Fig.~\ref{fig:3}, we also plot the variances $\overline{|(U_\text{nn})_{\alpha\beta}|^2}$ (solid lines) versus $\omega$ for the two sets of matrix elements considered. Comparing these variances makes apparent that they are qualitatively similar, but quantitatively different. The differences are best seen for $\omega \lesssim 5$. For $\omega \gtrsim 5$, both variances exhibit a similar exponential decay. Qualitatively similar results were obtained, not shown, for $\hat{U}_\text{n}$ and $\hat{K}_\text{nn}$.

\begin{figure}[!t]
\centering \includegraphics[width=\columnwidth]{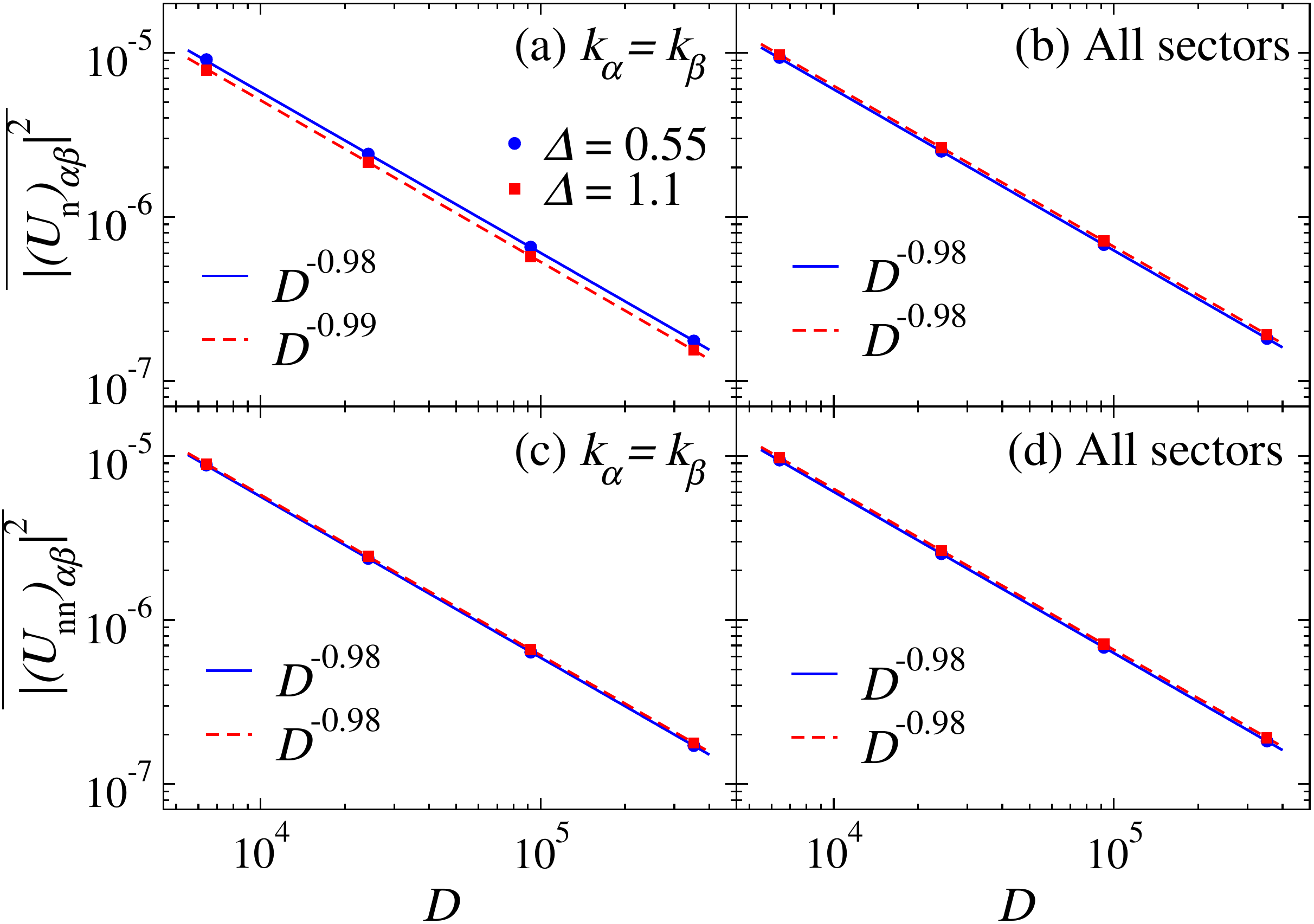}
\vspace{-0.5cm}
\caption{\label{fig:4} Scaling of $\overline{|(U_\text{n})_{\alpha\beta}|^2}$ [(a), (b)] and $\overline{|(U_\text{nn})_{\alpha\beta}|^2}$ [(c), (d)] vs $D$ at the nonintegrable ($\lambda=1$) point of Hamiltonian~\eqref{eq:xxz} with $\Delta=0.55$ and 1.1. We consider  pairs of energy eigenstates with $k_{\alpha}=k_{\beta}$ [(a), (c)] and pairs that mix all quasimomentum sectors [(b), (d)]. The straight lines show power-law fits to the results for $L=18$ through $L=22$. The average over $|O_{\alpha\beta}|^2$ for different chain sizes was calculated using pairs of energy eigenstates that satisfy $|\bar{E}|/L\leq 0.025$. We restricted the average to pairs of eigenstates for which $\omega<4$, the regime in which the variances exhibit a plateau-like behavior in Fig.~\ref{fig:3} (see Ref.~\cite{PRE2019} for scalings when one averages over all frequencies).} 
\end{figure}

Next, we study the scaling of the variances. Figure~\ref{fig:4} shows $\overline{|(U_\text{n})_{\alpha\beta}|^2}$ [(a), (b)] and $\overline{|(U_\text{nn})_{\alpha\beta}|^2}$ [(c), (d)] for $\Delta=0.55,\,1.1$ in chains with $L=16-22$. The averages are calculated over frequencies $\omega<4$ (qualitatively similar results were obtained averaging over other intervals of frequencies, see also Ref.~\cite{PRE2019}). The ETH ansatz~\eqref{eq:ETH} advances that the variances should scale as $1/D$ in the ``infinite-temperature'' regime, where $e^{S(\bar{E})}\simeq D$. The results in Fig.~\ref{fig:4} confirm that the variances for both observables and both sets of matrix elements (those for which $k_{\alpha}=k_{\beta}$ [(a), (c)] and those that connect all $k$-sectors [(b), (d)]) scale as $1/D$. In this respect, matrix elements of symmetry-breaking observables are no different than those of symmetry-preserving ones, despite the fact that the latter are nonvanishing only for $k_{\alpha}=k_{\beta}$. 

\subsection{Scaled Variances}\label{sec:svariancesQC}

The results in Fig.~\ref{fig:4} suggest that, for $\bar E\approx0$, one can define a Hilbert-space-size independent scaled variance
\beq\label{eq:scaledvarO}
|f_O(0,\omega)|^2= D\, \Var(O_{\alpha\beta}),
\eeq
as advanced by the ETH~\eqref{eq:ETH}. 

\begin{figure}[!b]
\centering \includegraphics[width=\columnwidth]{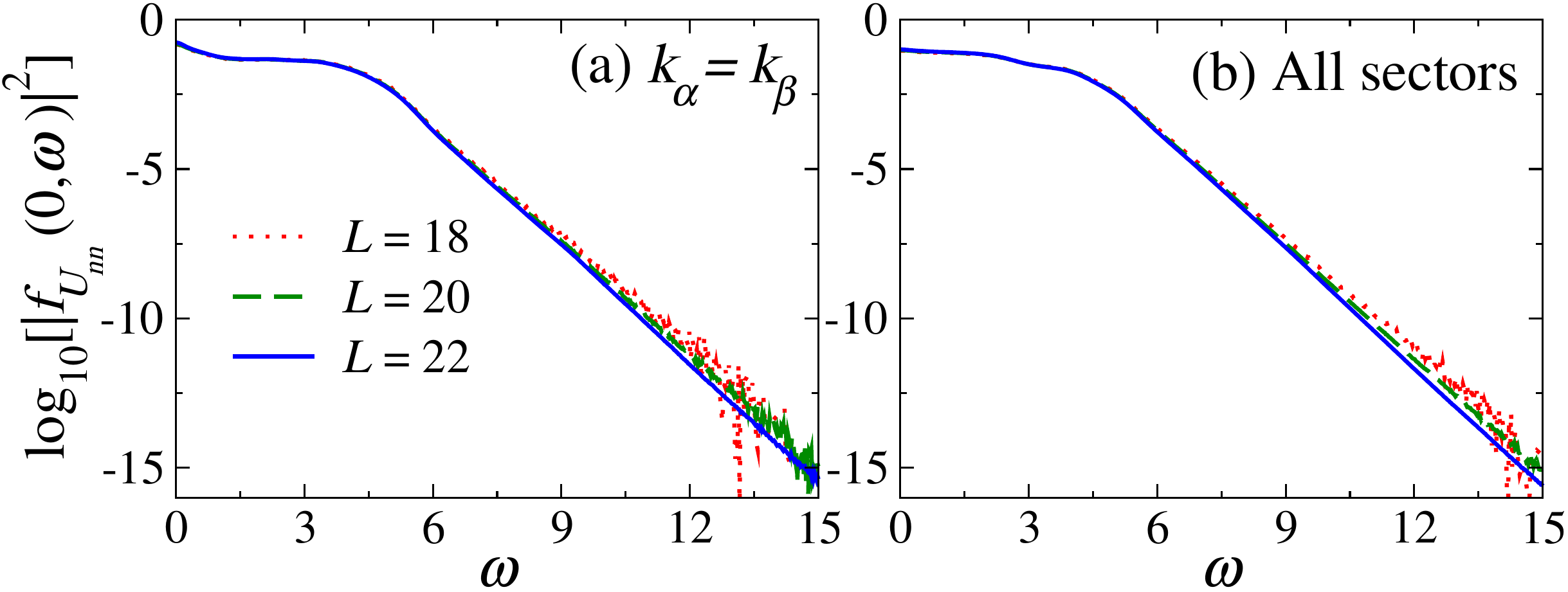}
\vspace{-0.4cm}
\caption{\label{fig:5} Scaled variance $|f_{U_\text{nn}}(0,\omega)|^2$ at the nonintegrable ($\lambda=1$) point of Hamiltonian~\eqref{eq:xxz} with $\Delta=0.55$ for different chain sizes $L$ (qualitatively similar results were obtained for $\Delta=1.1$). We show results for pairs of energy eigenstates with $k_{\alpha}=k_{\beta}$ (a) and pairs that mix all quasimomentum sectors (b). All pairs of eigenstates satisfy $|\bar{E}|/L\leq 0.025$. The averages $\overline{|(U_\text{nn})_{\alpha\beta}|^2}$ were coarse-grained in windows of width $\delta \omega = 0.025$.}
\end{figure}

In Fig.~\ref{fig:5}, we plot the scaled variance $|f_{U_\text{nn}}(0,\omega)|^2$ for three chain sizes. One can see that there is excellent data collapse away from the exponential regime at high $\omega$. In the latter regime, the scaled variances for contiguous chain sizes collapse over a larger $\omega$ window with increasing $L$. This points to finite-size effects as the reason for the lack of data collapse at high $\omega$. Larger finite-size effects are expected in finite chains at high frequencies because the matrix elements probe pairs of energy eigenstates at opposite edges of the energy spectrum~\cite{PRE2019}. Qualitatively similar results were found for all three observables studied irrespective of the Hamiltonian parameter $\Delta$. Altogether, our calculations show that for symmetry-breaking observables the function $|f_{O}(0,\omega)|^2$ is a well-defined smooth function of $\omega$.

We note that, for translationally invariant intensive observables such as the ones in Eqs.~\eqref{eq:UNNT}--\eqref{eq:KNNNT}, which have a Hilbert-Schmidt norm that scales as $1/\sqrt{L}$, the scaled variance was computed in Ref.~\cite{PRE2019} as
\beq\label{eq:scaledvarOT}
|f^T_O(0,\omega)|^2= {\mathcal D}L\, \Var(O^T_{\alpha\beta}),
\eeq
where $\mathcal D$ was the dimension of the specific symmetry sector considered. The results from Eq.~\eqref{eq:scaledvarOT} are consistent with the results from Eq.~\eqref{eq:scaledvarO} when one restricts the variance in the latter to only include pairs of states with $k_\alpha=k_\beta$. This is the case because, for $k_\alpha=k_\beta$, $\Var(O_{\alpha\beta})=\Var(O^T_{\alpha\beta})$ and $D\simeq {\mathcal D}L$.

\subsection{Low-Frequency Scaling}

\begin{figure}[!b]
\centering \includegraphics[width=\columnwidth]{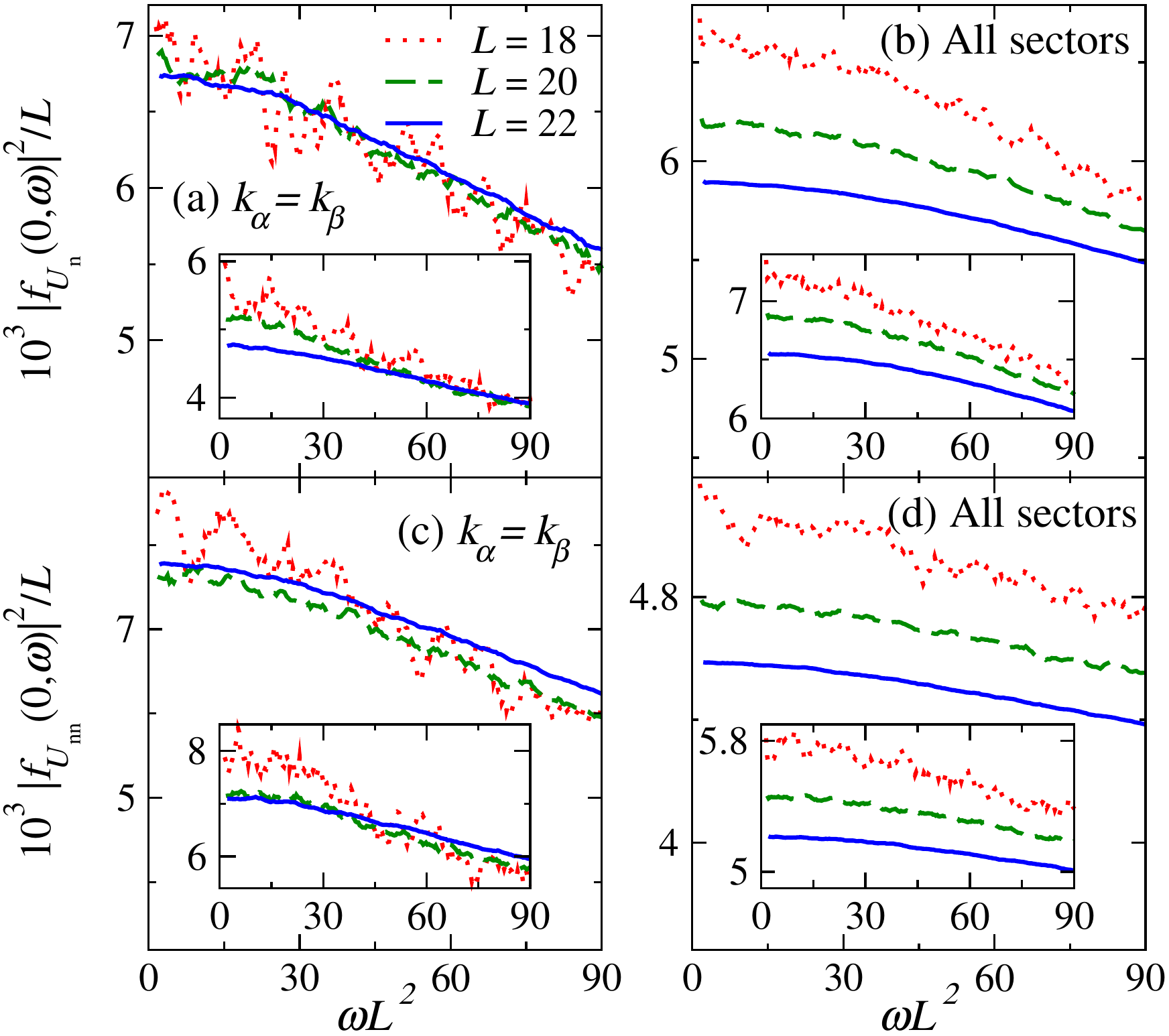}
\vspace{-0.4cm}
\caption{\label{fig:6} Low-frequency plots of the scaled variances $|f_O(0,\omega)|^2/L$ vs $\omega L^2$ for observables $\hat{U}_\text{n}$ [(a), (b)] and $\hat{U}_\text{nn}$ [(c), (d)] at the nonintegrable ($\lambda=1$) point of Hamiltonian~\eqref{eq:xxz}, with $\Delta=0.55$ (main panels) and $1.1$ (insets), for different chain sizes $L$. We consider pairs of energy eigenstates with $k_{\alpha}=k_{\beta}$ [(a), (c)] and pairs that mix all quasimomentum sectors [(b), (d)]. All pairs of eigenstates satisfy $|\bar{E}|/L\leq 0.025$. The running averages $\overline{|O_{\alpha\beta}|^2}$ were calculated in windows of width $\delta \omega = 0.009$ centered at points separated by $\Delta \omega=0.001$.}  
\end{figure}

For local operators in quantum chaotic systems, because of diffusion, one expects all dynamics to occur within times that scale with $L^2$. In the frequency domain, this means that $|f_{O}(\bar{E},\omega)|^2$ is expected to exhibit a plateau as $\omega\rightarrow0$ whose size (which defines the so-called Thouless energy) scales as $1/L^2$. Below the Thouless energy, the ETH ansatz coincides with the (featureless) predictions of random matrix theory. The magnitude of $|f_{O}(\bar{E},\omega)|^2$ in the plateau is expected to be proportional to $L$~\cite{ETH_review}. Such expectations have been confirmed in lattice systems with no translational symmetry (but no disorder)~\cite{ETH_review, brenes2020ballistic}, and the plateau has also been observed and its size characterized in systems with weak disorder~\cite{Serbyn17}.

Next, we study the low-frequency behavior of $|f_{O}(\bar{E},\omega)|^2$ for translational symmetry-breaking and symmetry-preserving operators in the energy eigenstates of the translationally invariant Hamiltonian~\eqref{eq:xxz} with $\lambda=1$  (in the quantum-chaotic regime). 

In Fig.~\ref{fig:6}, we plot $|f_O(0,\omega)|^2/L$ versus $\omega L^2$ for $\hat{U}_\text{n}$ [(a), (b)] and $\hat{U}_\text{nn}$ [(c), (d)] using pairs of energy eigenstates with $k_\alpha=k_\beta$ [(a), (c)] and pairs that connect all quasimomentum sectors [(b), (d)]. The main panels (insets) show results for $\Delta=0.55$ ($\Delta=1.1$). All the results reported in Fig.~\ref{fig:6} are consistent with the function $|f_O(0,\omega L^2)|^2/L$ becoming system-size independent for large systems at low $\omega$. Namely, they are consistent with the scaling advanced for quantum chaotic systems~\cite{ETH_review}. From Fig.~\ref{fig:6}, given the finite-size effects, it remains a challenge to extract the Thouless energy. 

\begin{figure}[!t]
\centering \includegraphics[width=\columnwidth]{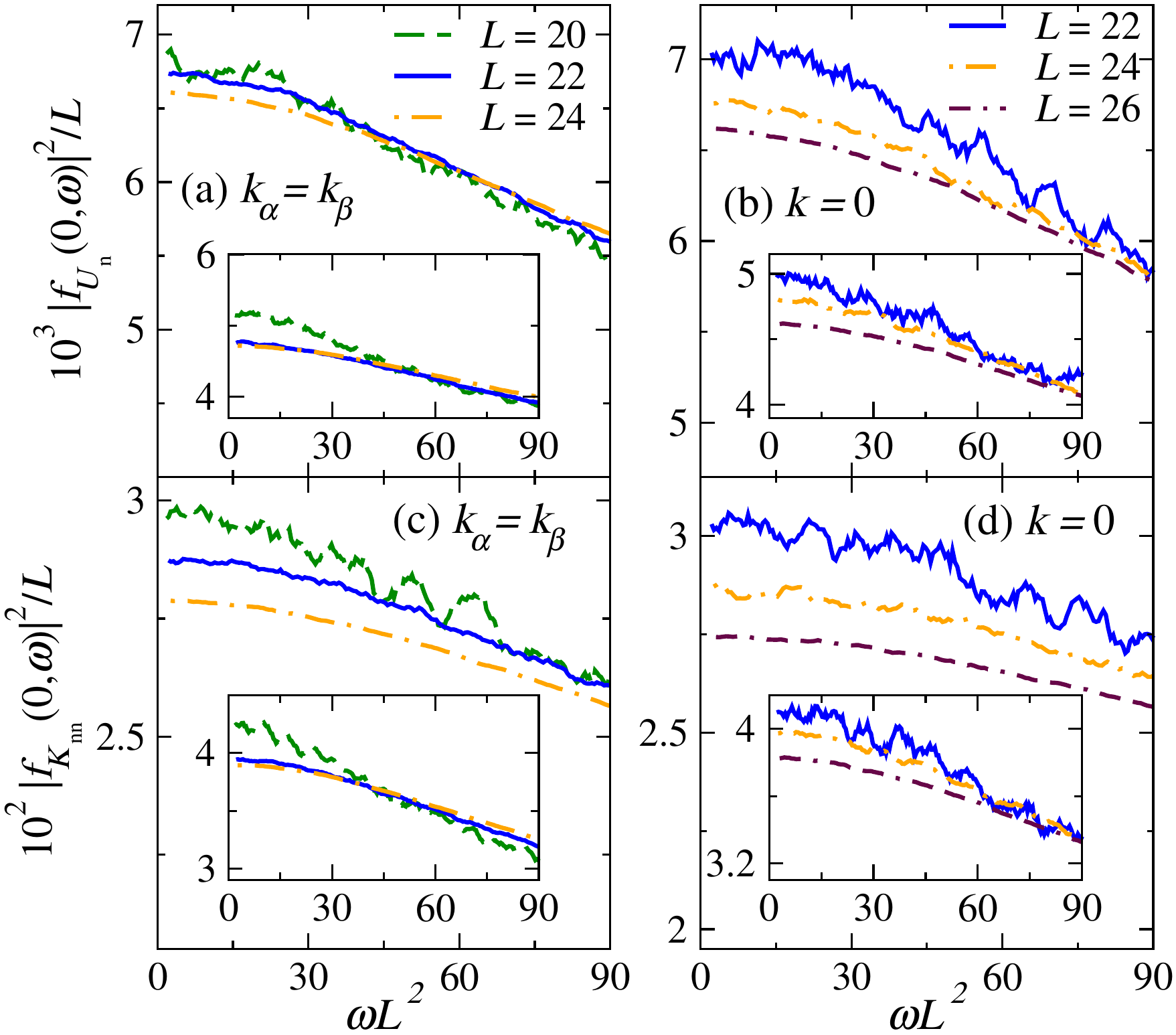}
\vspace{-0.4cm}
\caption{\label{fig:7} Low-frequency plots of the scaled variances $|f_O(0,\omega)|^2/L$ vs $\omega L^2$ for observables $\hat{U}_\text{n}$ (a) and $\hat{K}_\text{nn}$ (c), and of $|f^T_O(0,\omega)|^2/L$ for observables $\hat{U}^T_\text{n}$ (b) and $\hat{K}^T_\text{nn}$ (d), at the nonintegrable ($\lambda=1$) point of Hamiltonian~\eqref{eq:xxz}, with $\Delta=0.55$ (main panels) and $1.1$ (insets), for different chain sizes $L$. We consider pairs of energy eigenstates with $k_{\alpha}=k_{\beta}$ [(a), (c)] and within the even-$Z_2$, even-$P$ subsector of the $k=0$ sector [(b), (d)]. All pairs of eigenstates satisfy $|\bar{E}|/L\leq 0.025$. The running averages $\overline{|O_{\alpha\beta}|^2}$ were calculated in windows of width $\delta \omega = 0.009$ centered at points separated by $\Delta \omega=0.001$.} 
\end{figure}

Since the results in Fig.~\ref{fig:6} for pairs of energy eigenstates with $k_\alpha=k_\beta$ [(a), (c)] are qualitatively similar to those of pairs that connect all quasimomentum sectors [(b), (d)], albeit with smaller finite-size effects in the former (i.e., for symmetry-preserving observables) than in the latter (i.e., for symmetry-breaking observables), we focus on symmetry-preserving observables next. In Figs.~\ref{fig:7}(a) and~\ref{fig:7}(c), we plot $|f_O(0,\omega)|^2/L$ versus $\omega L^2$ for $\hat{U}_\text{n}$ and $\hat{K}_\text{nn}$, respectively, in pairs of energy eigenstates with $k_\alpha=k_\beta$ for chains with up to $L=24$, for $\Delta=0.55$ (main panels) and for $\Delta=1.1$ (insets). The agreement between the results for $\Delta=1.1$ (insets) in the two largest chains is much better than in Fig.~\ref{fig:6} [finite-size effects remain large for $\Delta=0.55$ (main panels)]. The results in Figs.~\ref{fig:7}(a) and~\ref{fig:7}(c) further strengthen the expectation that the function $|f_O(0,\omega L^2)|^2/L$ becomes, at low $\omega$, system-size independent for large systems. 

In Figs.~\ref{fig:7}(b) and~\ref{fig:7}(d), we plot $|f^T_O(0,\omega)|^2/L$ versus $\omega L^2$ for $\hat{U}^T_\text{n}$ and $\hat{K}^T_\text{nn}$, in the even-$Z_2$, even-$P$ subsector of the $k=0$ sector for chains with up to $L=26$, for $\Delta=0.55$ (main panels) and for $\Delta=1.1$ (insets). These are low-frequency results corresponding to the scaled variances reported in Ref.~\cite{PRE2019} for intermediate and large values of $\omega$. Figures~\ref{fig:7}(b) and~\ref{fig:7}(d) show that the behavior in the $k=0$ sector is qualitatively similar to the behavior for all pairs of energy eigenstates with $k_\alpha=k_\beta$ [Figs.~\ref{fig:7}(a) and~\ref{fig:7}(c)], but exhibits stronger finite-size effects. This suggests that, in exact diagonalization studies of matrix elements of translationally invariant operators, it may be better (in terms of reducing finite-size effects) to study averages over all quasimomentum sectors (excluding the $k=0$ and $\pi$ sectors) in smaller chains than to focus on the $k=0$ sector in larger ones.

\section{Interacting Integrable Chain} \label{integ}

Next, for the interacting integrable XXZ chain [$\lambda=0$ in Hamiltonian~\eqref{eq:xxz}], we carry out an analysis parallel to the one in the previous section. We show that the key results of Ref.~\cite{PRE2019} remain valid for symmetry-breaking observables, including a skewed log-normal-like distribution of off-diagonal matrix elements and a variance that is a smooth function of $\omega$ that scales as $1/D$. Additionally, we extend the analysis of Ref.~\cite{PRE2019} by identifying low-frequency ballistic and diffusive scalings of the variance of the off-diagonal matrix elements of both symmetry-breaking and symmetry-preserving observables. Lastly, we highlight differences between integrability-breaking and integrability-preserving observables, supporting the findings of Refs.~\cite{p2020adiabatic, brenes2020ballistic}.

\subsection{Distributions}

\begin{figure*}[!t]
\centering
\includegraphics[width=0.9\textwidth]{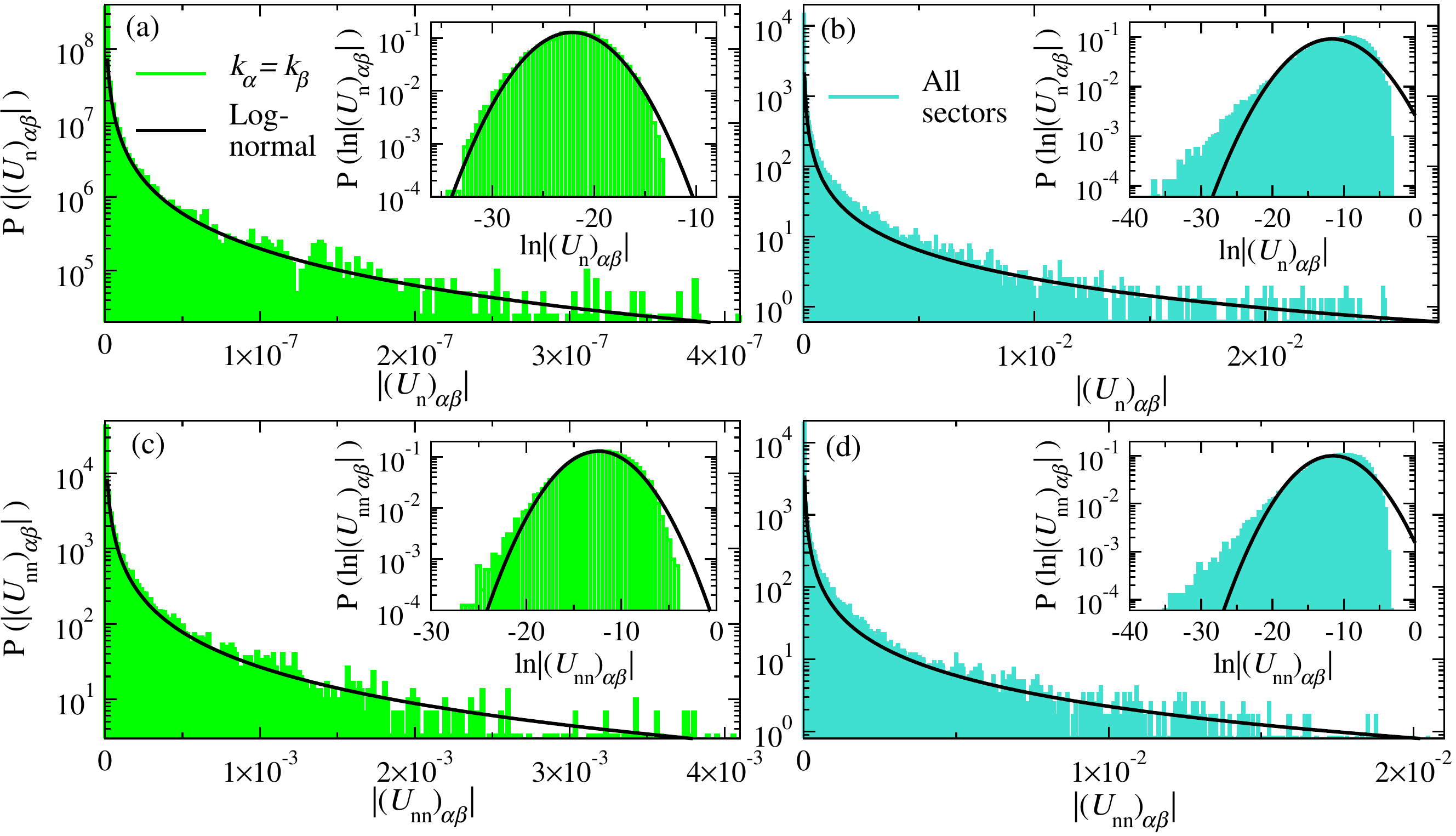}
\vspace{-0.2cm}
\caption{\label{fig:8} Probability distributions $P(|O_{\alpha\beta}|)$ for observables $\hat{U}_\text{n}$ [(a), (b)] and $\hat{U}_\text{nn}$ [(c), (d)] for Hamiltonian~\eqref{eq:xxz} with $\Delta=0.55$ (similar results were obtained for $\Delta=1.1$) and $\lambda=0$ (the integrable XXZ chain). We consider pairs of energy eigenstates for which $|\bar{E}|/L \leq 0.025$, and choose the 40,000 matrix elements with the lowest $\omega$ (this results in $\omega \leq 0.001$). We show results for matrix elements with $k_{\alpha}=k_{\beta}$ [(a), (c)] and matrix elements that mix all quasimomentum sectors [(b), (d)] in the $L=22$ chain. The insets show the probability distributions $P(\ln{|O_{\alpha\beta}|})$, along with Gaussian distributions (continuous lines) with the same mean and variance. The continuous lines in the main panels are the corresponding log-normal distributions.} 
\end{figure*}

Figure~\ref{fig:8} shows the distributions of $|O_{\alpha\beta}|$ for $\hat{U}_\text{n}$ [(a), (b)] and $\hat{U}_\text{nn}$ [(c), (d)] for matrix elements for which $\bar{E}\approx 0$ and $\omega \approx 0$. One can see that, regardless of whether matrix elements connect pairs of eigenstates from the same quasimomentum sectors [(a), (c)] or from all sectors [(b), (d)], the distributions are close to log-normal (the solid black lines are log-normal distributions with the same mean and variance as $\ln{|O_{\alpha\beta}|}$). Qualitatively similar results were obtained (not shown) for other frequencies, and for $\hat{K}_\text{nn}$.

A closer inspection of the distributions of $\ln{|O_{\alpha\beta}|}$ (insets) reveals the nature of the differences between the $P(|O_{\alpha\beta}|)$ and log-normal distributions. Specifically, the insets show that the $\ln{|O_{\alpha\beta}|}$ distributions are skewed normal, with a skewness that depends both on the observable [compare the insets in Figs.~\ref{fig:8}(a) and~\ref{fig:8}(c)] and on whether one looks at matrix elements that connect energy eigenstates from the same [Figs.~\ref{fig:8}(a) and~\ref{fig:8}(c)] or from all [Figs.~\ref{fig:8}(b) and~\ref{fig:8}(d)] quasimomentum sectors. For the three observables and the two values of $\Delta$ ($\Delta=0.55$ and 1.1) studied, we found that the distributions of matrix elements involving eigenstates from all quasimomentum sectors are the ones that exhibit a higher skewness. In Appendix~\ref{appendixskew}, we report a preliminary analysis that suggests that the distributions are skewed log-normal-like in the thermodynamic limit.

\subsection{Variances}

The lack of normality in the distribution of off-diagonal matrix elements of observables in integrable models means that the variance of the distribution does not determine other moments. Thus, there is no meaningful equivalent of the off-diagonal part of the ETH~\eqref{eq:ETH} in integrable systems. Still, the variance $\Var(O_{\alpha\beta})=\overline{|O_{\alpha\beta}|^2}$ (because $\overline{O_{\alpha\beta}}=0$) is what is physically relevant, e.g., for fluctuation-dissipation relations~\cite{ETH_review, Khatami}, heating rates~\cite{heatingrates}, transport properties~\cite{Steinigeweg2013Eigenstate, brenes2020eigenstate}, and the multipartite entanglement structure of energy eigenstates~\cite{Brenes2}. Thus, next, we seek to characterize the variance of the distribution of off-diagonal elements for symmetry-breaking observables and compare it to that of symmetry-preserving ones in the integrable XXZ chain.

\begin{figure}[!t]
\centering \includegraphics[width=\columnwidth]{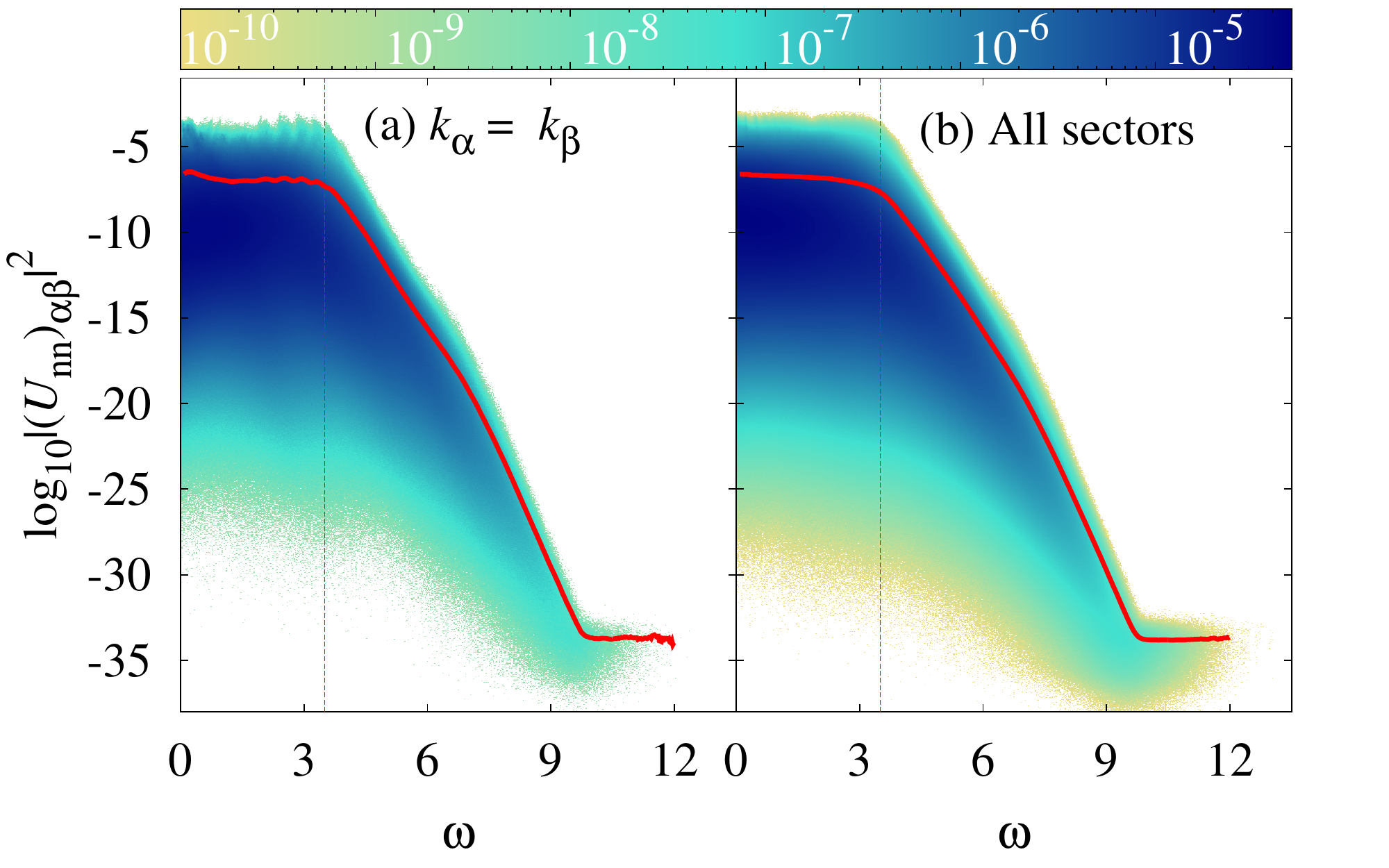}
\vspace{-0.6cm}
\caption{\label{fig:9} Normalized 2D histograms of $\log_{10}|(U_\text{nn})_{\alpha\beta}|^2$ vs $\omega$ in the XXZ chain with $\Delta=0.55$ for $L=22$ (qualitatively similar results were obtained for $\Delta=1.1$). We consider pairs of energy eigenstates with $k_{\alpha}=k_{\beta}$ (a) and pairs that mix all quasimomentum sectors (b). All pairs of energy eigenstates satisfy $|\bar{E}|/L\le0.025$. The (red) solid lines are running averages $\log_{10}\overline{|(U_\text{nn})_{\alpha\beta}|^2}$ calculated in windows of width $\delta\omega=0.175$ centered at points separated by $\Delta\omega=0.025$. The vertical dashed lines show the values of $\omega$ up to which results for $|O_{\alpha\beta}|^2$ are included in the scaling analysis of Fig.~\ref{fig:10}.} 
\end{figure}

In Fig.~\ref{fig:9}, we show normalized 2D histograms of $\log_{10}{|(U_\text{nn})_{\alpha\beta}|^2}$ for pairs of energy eigenstates that satisfy $|\bar{E}|/L\leq 0.025$ in chains with $L=22$. We report results for $\Delta=0.55$ (the ones obtained for $\Delta=1.1$, not shown, are qualitatively similar) between pairs of eigenstates with $k_{\alpha}=k_{\beta}$ [Fig.~\ref{fig:9}(a)] and between pairs that connect all quasimomentum sectors [Fig.~\ref{fig:9}(b)]. We note that the results in Fig.~\ref{fig:9}(a) are qualitatively similar to those reported in Ref.~\cite{PRE2019} for translationally invariant observables in the $k=0$ sector. As in Ref.~\cite{PRE2019}, the support of the distribution for $\hat U_\text{nn}$ is much broader for the interacting integrable system [Fig.~\ref{fig:9}(a)] than for the nonintegrable one [Fig.~\ref{fig:3}(a)]. Also, in Fig.~\ref{fig:9}(a), no significant fraction of matrix elements has a vanishing magnitude as seen in quadratic models~\cite{Khatami}. Because of this, for interacting integrable models, one can define a meaningful average $\overline{|O_{\alpha\beta}|^2}$ at each value of $\omega$. Figure~\ref{fig:9}(b) shows that the same is true for symmetry-breaking observables that connect all quasimomentum sectors.

The (red) solid lines in Fig.~\ref{fig:9} show the $\omega$-resolved variances of $|(U_\text{nn})_{\alpha\beta}|$. As in the quantum chaotic case (Fig.~\ref{fig:3}), differences can be seen in the variances of matrix elements connecting the same quasimomentum sectors [Fig.~\ref{fig:9}(a)] and all sectors [Fig.~\ref{fig:9}(b)] for $\omega\lesssim 4$. The exponential and Gaussian regimes at high $\omega$ (see Ref.~\cite{PRE2019}) are similar in both sets of matrix elements.

\begin{figure}[!b]
\centering \includegraphics[width=\columnwidth]{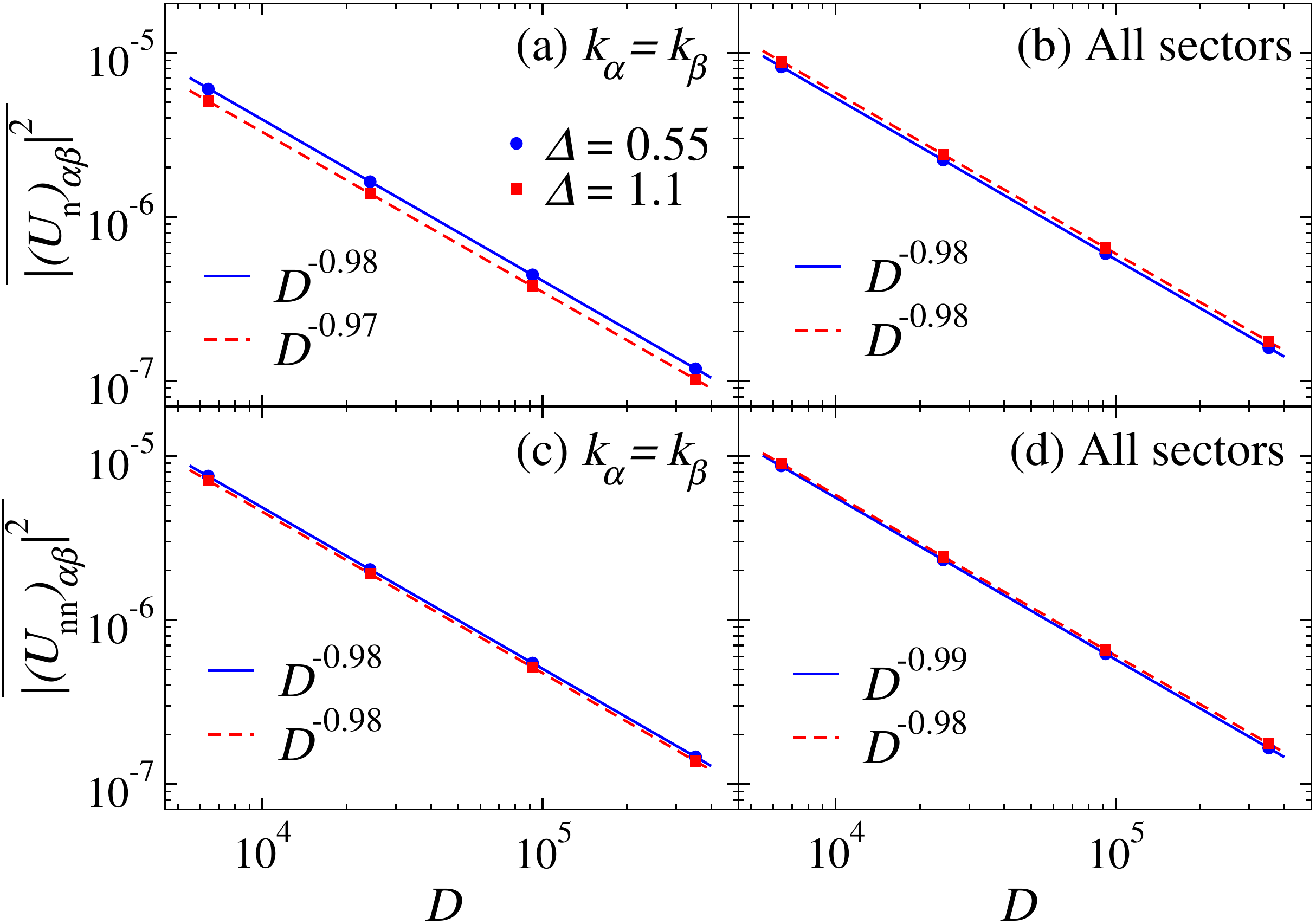}
\vspace{-0.4cm}
\caption{\label{fig:10} Scaling of $\overline{|(U_\text{n})_{\alpha\beta}|^2}$ [(a), (b)] and $\overline{|(U_\text{nn})_{\alpha\beta}|^2}$ [(c), (d)] vs $D$ in the (integrable) XXZ chain with $\Delta=0.55$ and 1.1. We consider pairs of energy eigenstates with $k_{\alpha}=k_{\beta}$ [(a), (c)] and pairs that mix all quasimomentum sectors [(b), (d)]. The straight lines show power-law fits to the results for $L=18$ through $L=22$. The average over $|O_{\alpha\beta}|^2$ for different chain sizes was calculated using pairs of energy eigenstates that satisfy $|\bar{E}|/L\leq 0.025$. We restricted the average to pairs of eigenstates for which $\omega<3.5$, the regime in Fig.~\ref{fig:9} in which the variances exhibit a plateau-like behavior (see Ref.~\cite{PRE2019} for scalings when one averages over all frequencies).}
\end{figure}

Next, we study how the variances scale with increasing chain size. In Fig.~\ref{fig:10}, we show finite-size scaling analyses of the variance $\overline{|O_{\alpha\beta}|^2}$ versus $D$ for $\hat{U}_\text{n}$ [(a), (b)] and $\hat{U}_\text{nn}$ [(c), (d)] for chains with $L=16-22$. The average is calculated over frequencies $\omega<3.5$ (qualitatively similar results were obtained averaging over other intervals of frequencies, see also Ref.~\cite{PRE2019}). As found in Ref.~\cite{PRE2019} for translationally invariant observables in the $k=0$ sector of the XXZ chain, all variances in Fig.~\ref{fig:10} scale as $1/D$ (as they do in the quantum chaotic system in Fig.~\ref{fig:4}). This occurs regardless of whether the matrix elements are computed between pairs of energy eigenstates from the same quasimomentum sector [(a), (c)] or between pairs that mix all quasimomentum sectors [(b), (d)]. 

\subsection{Scaled Variances}\label{sec:IsvariancesQC}

The results in Fig.~\ref{fig:10} suggest that, for $\bar E\approx0$, one can define a Hilbert-space-size independent scaled variance
\beq\label{eq:IscaledvarO}
V_O(0,\omega)= D\, \Var(O_{\alpha\beta}),
\eeq
as for quantum-chaotic systems~\eqref{eq:scaledvarO}. Note that we use a different label for the scaled variance in integrable systems to emphasize that there is no equivalent of the off-diagonal part of the ETH~\eqref{eq:ETH} for them.

\begin{figure}[!b]
\centering \includegraphics[width=\columnwidth]{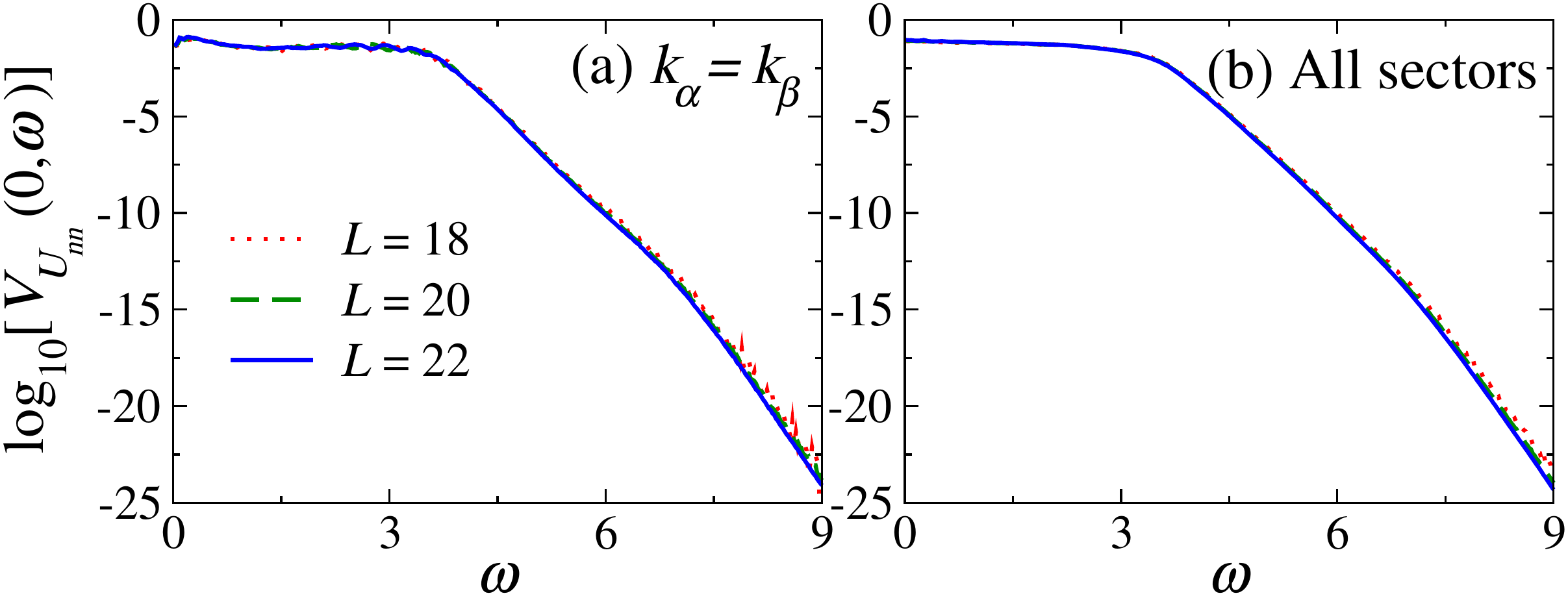}
\vspace{-0.4cm}
\caption{\label{fig:11} Scaled variance $V_{U_\text{nn}}(0,\omega)$ in the XXZ chain with $\Delta=0.55$ for different chain sizes $L$ (qualitatively similar results were obtained for $\Delta=1.1$). We show results for pairs of energy eigenstates with $k_{\alpha}=k_{\beta}$ (a) and pairs that mix all quasimomentum sectors (b). All pairs of eigenstates satisfy $|\bar{E}|/L\leq 0.025$. The averages $\overline{|(U_\text{nn})_{\alpha\beta}|^2}$ were coarse-grained in windows of width $\delta \omega = 0.025$.}
\end{figure}

In Fig.~\ref{fig:11}, we plot the scaled variance $V_{U_\text{nn}}(0,\omega)$ for three chain sizes. The results in Fig.~\ref{fig:11}(a) confirm the data collapse expected from Ref.~\cite{PRE2019} for symmetry-preserving observables, while the results in Fig.~\ref{fig:11}(b) demonstrate that the same is true for symmetry-breaking ones. We note that, for translationally invariant intensive observables such as the ones in Eqs.~\eqref{eq:UNNT}--\eqref{eq:KNNNT}, which have a Hilbert-Schmidt norm that scales as $1/\sqrt{L}$, the scaled variance was computed in Ref.~\cite{PRE2019} as
\beq\label{eq:IscaledvarOT}
V^T_O(0,\omega)= {\mathcal D}L\, \Var(O^T_{\alpha\beta}),
\eeq
where $\mathcal D$ was the dimension of the specific symmetry sector considered. The results from Eq.~\eqref{eq:IscaledvarOT} are consistent with the results from Eq.~\eqref{eq:IscaledvarO} for states with $k_\alpha=k_\beta$ because $\Var(O_{\alpha\beta})=\Var(O^T_{\alpha\beta})$ and $D\simeq {\mathcal D}L$.

In Fig.~\ref{fig:11}, finite-size effects are smaller for the smallest values of $V_{U_\text{nn}}(0,\omega)$ computed than in nonintegrable systems (see Fig.~\ref{fig:5}). The reason is that $V_{U_\text{nn}}(0,\omega)$ decays more quickly with $\omega$ in integrable systems~\cite{PRE2019, heatingrates} so that, for the smallest values of $V_{U_\text{nn}}(0,\omega)$ computed (limited by the machine precision) for the largest chains, the matrix elements are not probing the edges of the spectrum. 

Overall, the results in Fig.~\ref{fig:11} strengthen the conclusion in Ref.~\cite{PRE2019}, explored recently in nontranslationally invariant XXZ chains~\cite{brenes2020eigenstate, brenes2020ballistic}, that in interacting integrable systems there is a well defined scaled variance $V_O(\bar E,\omega)$. As per our results here, the scaled variance is well defined even for observables that break Hamiltonian symmetries.

\subsection{Low-Frequency Scaling}

\begin{figure}[!b]
\centering \includegraphics[width=\columnwidth]{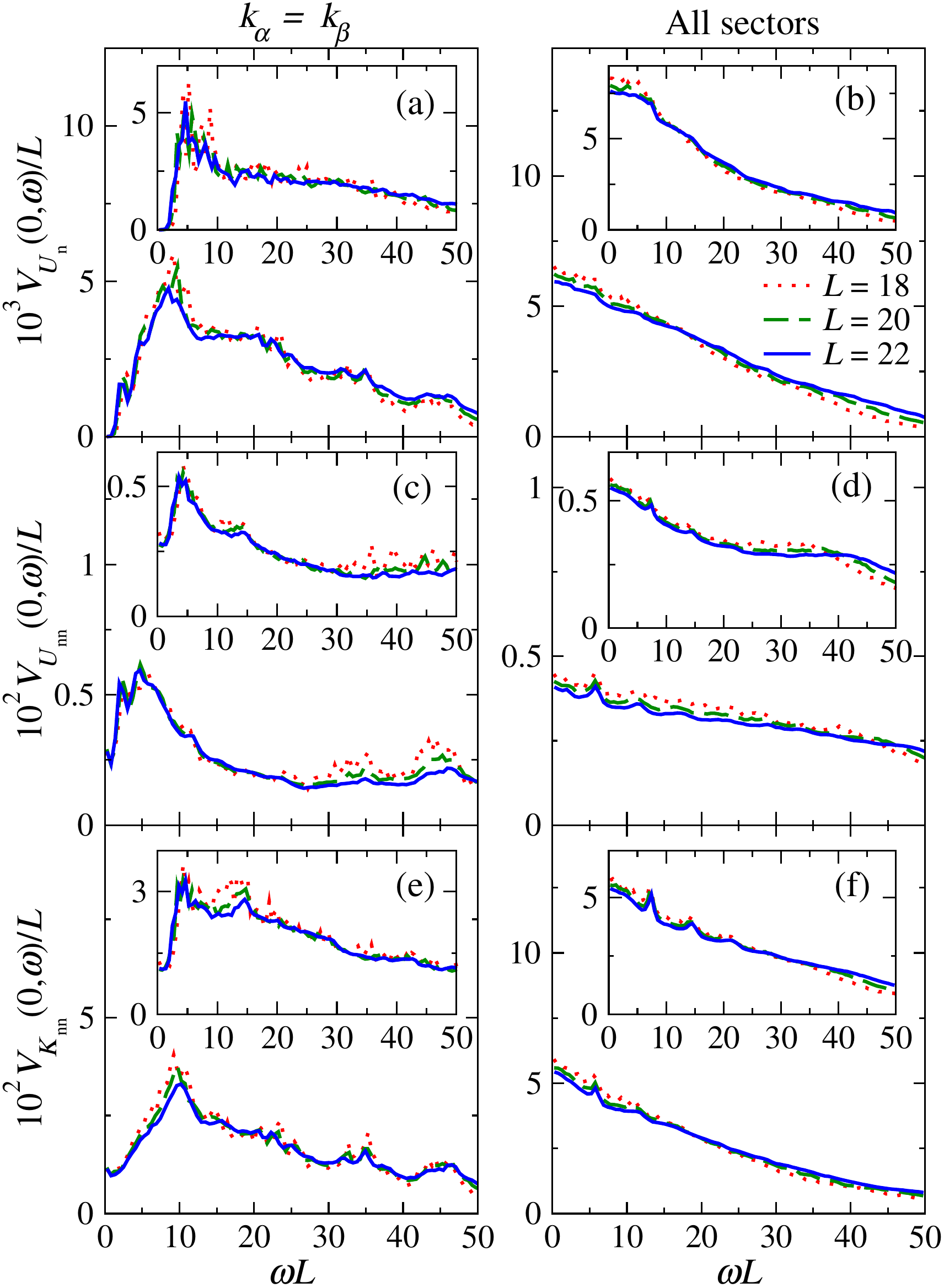}
\vspace{-0.4cm}
\caption{\label{fig:12} Low-frequency plots of the scaled variances $V_O(0,\omega)/L$ vs $\omega L$ for observables $\hat{U}_\text{n}$ [(a), (b)], $\hat{U}_\text{nn}$ [(c), (d)], and $\hat{K}_\text{nn}$ [(e), (f)] in the (integrable) XXZ chain with $\Delta=0.55$ (main panels) and 1.1 (insets), for different chain sizes $L$. We consider pairs of energy eigenstates with $k_{\alpha}=k_{\beta}$ [(a), (c), (e)] and pairs that mix all quasimomentum sectors [(b), (d), (f)]. All pairs of eigenstates satisfy $|\bar{E}|/L\leq 0.025$. The averages $\overline{|O_{\alpha\beta}|^2}$ were coarse-grained in windows of width $\delta \omega = 0.025$.}
\end{figure}

\begin{figure}[!b]
\centering \includegraphics[width=\columnwidth]{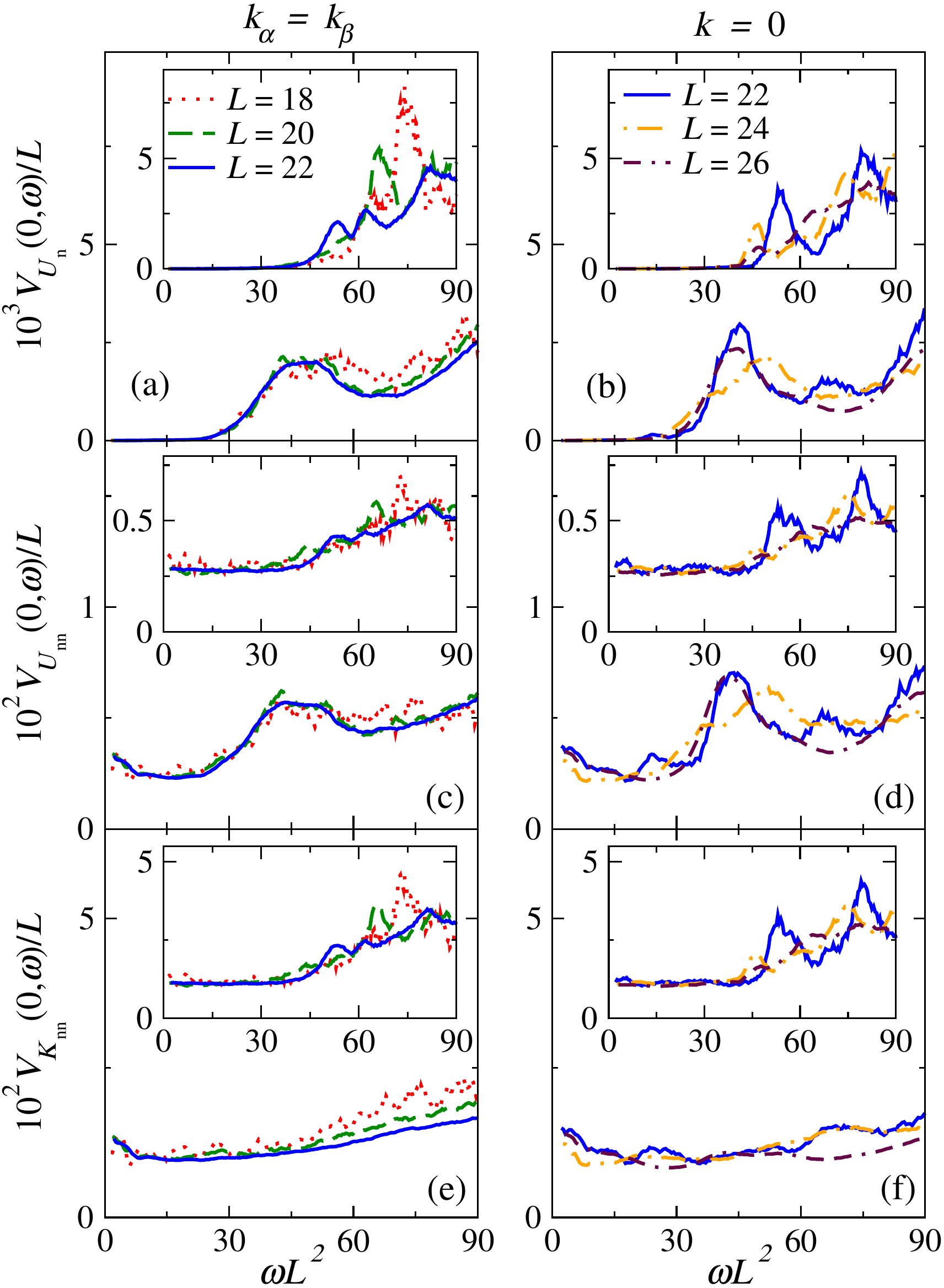}
\vspace{-0.4cm}
\caption{\label{fig:13} Low-frequency plots vs $\omega L^2$ of the scaled variances $V_O(0,\omega)/L$ for observables $\hat{U}_\text{n}$ (a), $\hat{U}_\text{nn}$ (c), $\hat{K}_\text{nn}$ (e), and of $V^T_O(0,\omega)/L$ for observables $\hat{U}^T_\text{n}$ (b), $\hat{U}^T_\text{nn}$ (d), and $\hat{K}^T_\text{nn}$ (f), in the XXZ chain with $\Delta=0.55$ (main panels) and 1.1 (insets), for different chain sizes $L$. We consider pairs of energy eigenstates with $k_{\alpha}=k_{\beta}$ [(a), (c), (e)] and within the even-$Z_2$, even-$P$ subsector of the $k=0$ sector [(b), (d), (f)]. All pairs of eigenstates satisfy $|\bar{E}|/L\leq 0.025$. The running averages $\overline{|O_{\alpha\beta}|^2}$ were calculated in windows of width $\delta \omega = 0.009$ centered at points separated by $\Delta \omega=0.001$.}  
\end{figure}

\begin{figure*}[!t]
\centering
\includegraphics[width=0.93\textwidth]{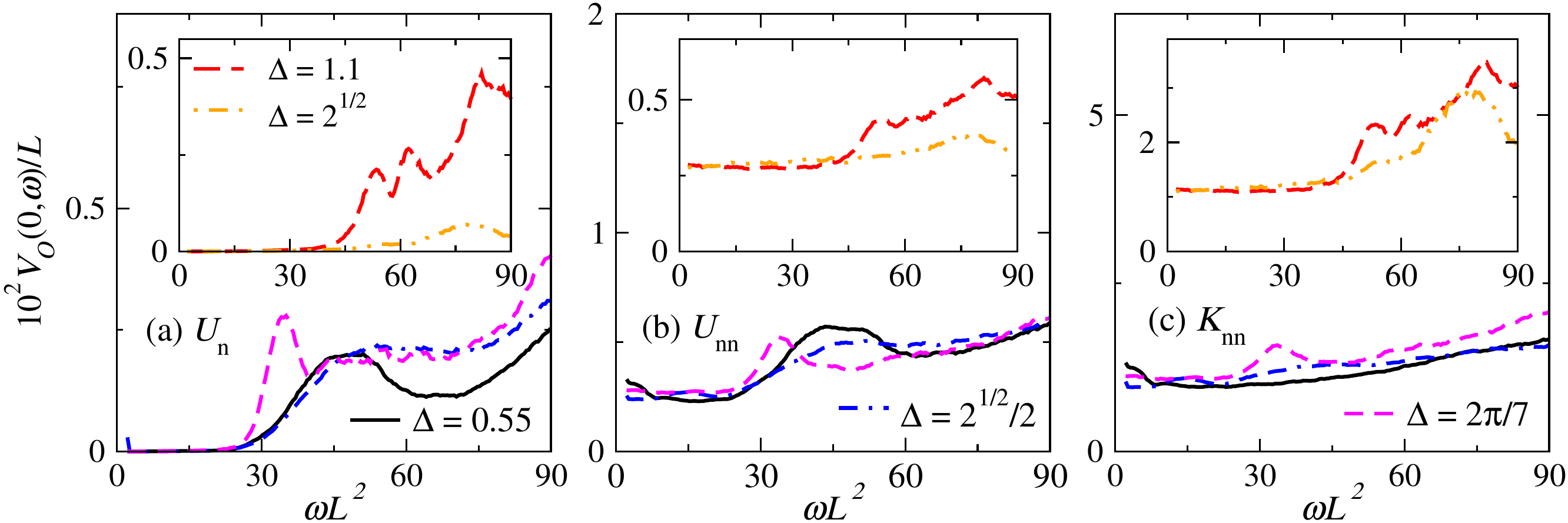}
\vspace{-0.2cm}
\caption{\label{fig:dfig} Low-frequency plots of the scaled variances $V_O(0,\omega)/L$ vs $\omega L^2$ for observables $\hat{U}_\text{n}$ (a), $\hat{U}_\text{nn}$ (b), and $\hat{K}_\text{nn}$ (c) in the XXZ chain with $\Delta$ in the easy-plane regime ($\Delta<1$, main panels) and the easy-axis regime ($\Delta>1$, insets). We consider pairs of energy eigenstates with $k_{\alpha}=k_{\beta}$. All pairs of eigenstates satisfy $|\bar{E}|/L\leq 0.025$. The running averages $\overline{|O_{\alpha\beta}|^2}$ were calculated in windows of width $\delta \omega = 0.009$ centered at points separated by $\Delta \omega=0.001$.} 
\end{figure*}

Next we study the low-frequency behavior of the scaled variances $V_O(0,\omega)$. Two recent works~\cite{brenes2020eigenstate, brenes2020ballistic} have studied the low-frequency behavior of scaled variances of nontranslationally invariant operators like the ones in Eqs.~\eqref{eq:obs_Unn}--\eqref{eq:obs_Knnn}, and of averages like the ones in Eqs.~\eqref{eq:UNNT}--\eqref{eq:KNNNT}, in the XXZ chain with open boundary conditions (namely, without translational symmetry). For the average spin current operator per site, in Ref.~\cite{brenes2020eigenstate} it was shown that the scaled variance exhibits a large low-frequency peak in the easy-plane regime ($\Delta=0.55$) whose height is proportional to $L$ and location in frequency scales as $1/L$. The area under the peak does not change with increasing system size, and in the thermodynamic limit it is expected to signal ballistic DC transport (the peak would be at $\omega=0$ and it would have a nonzero weight)~\cite{rigol_shastry_08, marlon2018}. Such a peak was absent in the scaled variance in the easy-axis ($\Delta=1.1$) regime~\cite{brenes2020eigenstate}. For other observables, the results in Ref.~\cite{brenes2020ballistic} are qualitatively similar to results that we report here so we will mention them along with our discussion.

In Fig.~\ref{fig:12}, we plot $V_O(0,\omega)/L$ versus $\omega L$ in chains with up to $L=22$ for $\hat{U}_\text{n}$ [(a), (b)], $\hat{U}_\text{nn}$ [(c), (d)], and $\hat{K}_\text{nn}$ [(e), (f)]. In the left column [(a), (c), (e)], we show results for pairs of energy eigenstates from the same quasimomentum sectors and, in the right column [(b), (d), (f)], we show results for pairs that connect all quasimomentum sectors. In the main panels (insets), we show results for $\Delta=0.55$ ($\Delta=1.1$). All plots in Fig.~\ref{fig:12} exhibit good data collapse. In particular, one can see that the location of small features (e.g., peaks and valleys) does not change for different chain sizes (see also the results in Appendix~\ref{appendix}). This shows that in the XXZ chain, both in the easy-plane and easy-axis regimes, as well as for both symmetry-preserving and symmetry-breaking observables, there is a robust regime in which the variances $V_O(0,\omega)/L$ exhibit ballistic scalings. Qualitatively similar results were reported in Ref.~\cite{brenes2020ballistic} for the XXZ chain with open boundary conditions. Ballistic scalings of variances have also been observed in quantum-chaotic systems~\cite{ETH_review, brenes2020eigenstate}. The collapse of the scaled variances $V_O(0,\omega)/L$ when plotted versus $\omega L$ degrades as $\omega$ increases and one enters the $L$ independent regime depicted in Fig.~\ref{fig:11}. Characterizing the transition between these two regimes is an interesting problem that should be tackled in future works.

\begin{figure*}[!t]
\centering \includegraphics[width=0.93\textwidth]{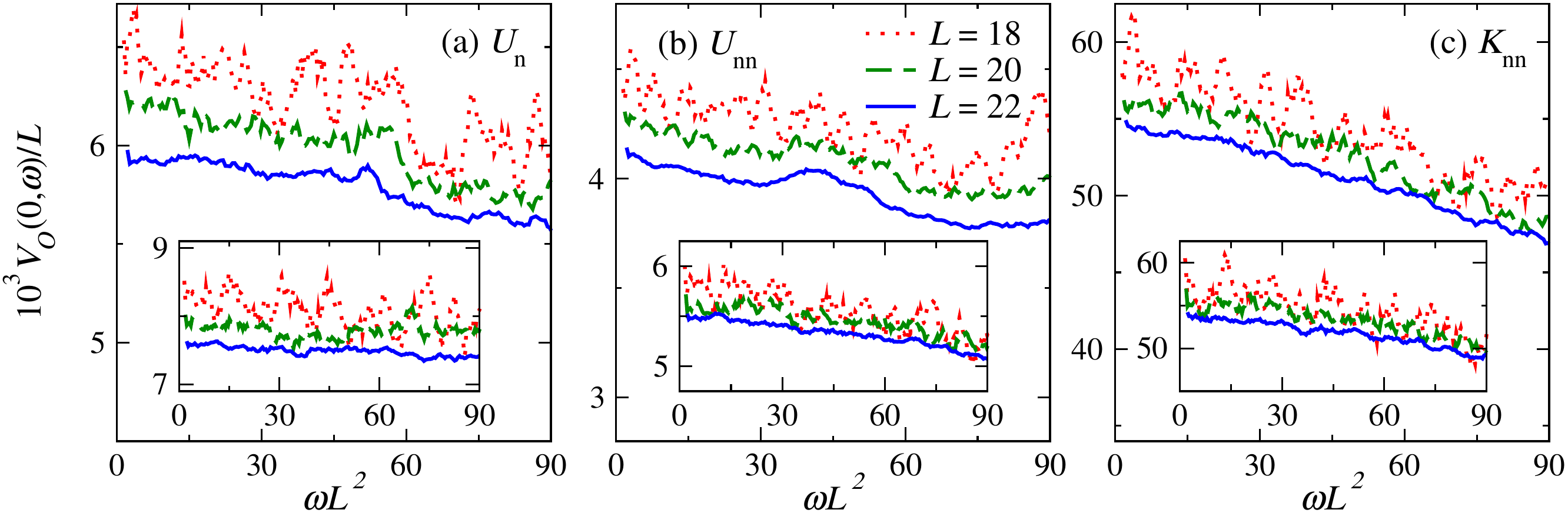}
\vspace{-0.2cm}
\caption{\label{fig:14} Low-frequency plots of the scaled variances $V_O(0,\omega)/L$ vs $\omega L^2$ for observables $\hat{U}_\text{n}$ (a), $\hat{U}_\text{nn}$ (b), and $\hat{K}_\text{nn}$ (c), in the XXZ chain with $\Delta=0.55$ (main panels) and 1.1 (insets), for different chain sizes $L$. We consider pairs of energy eigenstates that mix all quasimomentum sectors. All pairs of eigenstates satisfy $|\bar{E}|/L\leq 0.025$. The running averages $\overline{|O_{\alpha\beta}|^2}$ were calculated in windows of width $\delta \omega = 0.009$ centered at points separated by $\Delta \omega=0.001$.}
\end{figure*}

Let us focus first on the behavior of $V_O(0,\omega)/L$ for matrix elements that connect energy eigenstates from the same quasimomentum sectors (symmetry-preserving observables). Comparing the results in Fig.~\ref{fig:12}(a) with those in Figs.~\ref{fig:12}(c) and~\ref{fig:12}(e), one can see that $V_{{U}_\text{n}}(0,\omega\rightarrow0)/L$ vanishes while $V_{U_\text{nn}}(0,\omega\rightarrow0)/L$ and $V_{K_\text{nn}}(0,\omega\rightarrow0)/L$ converge to a nonzero system-size-independent value (see also the results in Appendix~\ref{appendix}). This behavior is qualitatively similar to the one reported in Ref.~\cite{brenes2020ballistic} for the XXZ chain with open boundary conditions. There, the scaled variance was found to vanish as $\omega\rightarrow0$ for observables that do not break the integrability of the XXZ chain (as is the case here for $\hat{U}^T_\text{n}$), while $V_O(0,\omega)/L$ was found to converge to a nonzero system-size-independent value for observables that do (as is the case here for $\hat U^T_\text{nn}$ and $\hat K^T_\text{nn}$). This is consistent with the results from Ref.~\cite{p2020adiabatic} for frequencies that are exponentially small in system size. However, we must emphasize that the results in Fig.~\ref{fig:12} and in Ref.~\cite{brenes2020ballistic} are for frequencies that are polynomially small in system size and, as such, involve an average over a rapidly (exponentially) growing number of matrix elements with increasing system size.

An interesting feature in the behavior of $V_O(0,\omega)/L$ in Figs.~\ref{fig:12}(a),~\ref{fig:12}(c), and~\ref{fig:12}(e), for both $\Delta=0.55$ (main panels) and 1.1 (insets), is that there is a worsening of the data collapse as $\omega\rightarrow0$ (it is difficult to see in the plots because it occurs at small values of $\omega L$). This was also noticed in results reported in Ref.~\cite{brenes2020ballistic}. In Figs.~\ref{fig:13}(a),~\ref{fig:13}(c), and~\ref{fig:13}(e), we replot (using a finer coarse graining) the lowest frequency results from Figs.~\ref{fig:12}(a),~\ref{fig:12}(c), and~\ref{fig:12}(e) but against $\omega L^2$. The excellent data collapse in Figs.~\ref{fig:13}(a),~\ref{fig:13}(c), and~\ref{fig:13}(e) at the lowest frequencies (see also the results in Appendix~\ref{appendix}) suggests that, no matter whether the XXZ chain is in the easy-plane or easy-axis regimes, the variances exhibit diffusive scalings. For completeness, in Figs.~\ref{fig:13}(b),~\ref{fig:13}(d), and~\ref{fig:13}(f), we plot $V^T_O(0,\omega)/L$ versus $\omega L^2$ for $\hat{U}^T_\text{n}$, $\hat{U}^T_\text{nn}$, and $\hat{K}^T_\text{nn}$ in the even-$Z_2$, even-$P$ subsector of the $k=0$ sector for chains with up to $L=26$ for both $\Delta=0.55$ (main panels) and $\Delta=1.1$ (insets). The results resemble the ones from Figs.~\ref{fig:13}(a),~\ref{fig:13}(c), and~\ref{fig:13}(e), but exhibit larger finite-size effects, as found in Fig.~\ref{fig:7} for quantum-chaotic systems.

To further explore the role of $\Delta$ in the low-frequency behavior of the scaled variances of symmetry preserving operators, in Fig.~\ref{fig:dfig} we plot $V_{{U}_\text{n}}(0,\omega)/L$ (a), $V_{{U}_\text{nn}}(0,\omega)/L$ (b), and $V_{{K}_\text{nn}}(0,\omega)/L$ (c) versus $\omega L^2$ for different values of the anisotropy parameter $\Delta$ for lattices with $L=22$~\footnote{Note that the new values of $\Delta$ considered are irrational numbers. Irrational values of $\Delta$ help eliminate non-generic features that may appear in the matrix elements because of finite-size effects.}. The main panels show results in the easy-plane regime, while the insets show results in the easy-axis regime. For $\hat{U}_\text{n}$, which is the integrability-preserving observable, $V_{{U}_\text{n}}(0,\omega\rightarrow0)/L$ vanishes irrespective of $\Delta$. Conversely, for the integrability-breaking observables $\hat{U}_\text{nn}$ and $\hat{K}_\text{nn}$, $V_{{U}_\text{nn}}(0,\omega\rightarrow0)/L$ and $V_{{K}_\text{nn}}(0,\omega\rightarrow0)/L$ do not vanish for any $\Delta$. In the lowest frequency regime for the latter observables, a robust plateau is seen in the scaled variances for $\Delta>1$, and the results for $\Delta<1$ are consistent with a plateau. Hence, our results suggest that, as in quantum chaotic systems, diffusion puts the ultimate limit on the equilibration time for integrability-breaking observables in the XXZ chain. 

To conclude, let us discuss the behavior of the variances for the operators that break translational symmetry. In Figs.~\ref{fig:12}(b),~\ref{fig:12}(d), and~\ref{fig:12}(f), we show results for $V_O(0,\omega)/L$ versus $\omega L$ when averaging over all matrix elements (i.e., for the symmetry-breaking operators). The scaled variances for the three observables, for $\Delta=0.55$ (main panels) and 1.1 (insets), are all qualitatively similar. The contrast with the results in Figs.~\ref{fig:12}(a),~\ref{fig:12}(c), and~\ref{fig:12}(e) for matrix elements within the same quasimomentum sectors (symmetry-preserving operators) is remarkable. Breaking translational symmetry does not affect the ballistic scaling of the variances but erases many features in $V_O(0,\omega)/L$, especially the vanishing [Fig.~\ref{fig:12}(a)] or the fast decrease [Figs.~\ref{fig:12}(c), and~\ref{fig:12}(e)] seen in $V_O(0,\omega)/L$ as $\omega\rightarrow0$. For all results in Figs.~\ref{fig:12}(b),~\ref{fig:12}(d), and~\ref{fig:12}(f), $V_O(0,\omega\rightarrow0)/L$ is seen to plateau to a (close to) system-size-independent value. Since $\hat{U}_\text{n}$, $\hat{U}_\text{nn}$, and $\hat{K}_\text{nn}$ break the integrability of the XXZ chain if added as perturbations, the observed behavior is consistent with our previous discussion for integrability-breaking observables. In Figs.~\ref{fig:14}(a),~\ref{fig:14}(b), and~\ref{fig:14}(c), we replot (using a finer coarse-graining) the lowest frequency results from Figs.~\ref{fig:12}(b),~\ref{fig:12}(d), and~\ref{fig:12}(f), respectively, but against $\omega L^2$. They are qualitatively similar to the results shown in Fig.~\ref{fig:6} for quantum chaotic systems. As in Fig.~\ref{fig:6}, larger finite-size effects for the symmetry-breaking observables appear to disrupt the expected scaling for the magnitude of the variance.

\section{Summary and Discussion} \label{conc}

We studied the off-diagonal matrix elements of observables that break translational symmetry in eigenstates of translationally invariant Hamiltonians. In contrast to translationally invariant observables, the matrix elements of the observables considered here connect sectors with different total quasimomentum. We probed properties of the matrix elements in a quantum-chaotic Hamiltonian, as well as in an interacting integrable one (the XXZ chain).

In the quantum-chaotic model, we found that the qualitative behavior of the off-diagonal matrix elements is unaffected by the block diagonal structure of the Hamiltonian in quasimomentum space. Namely, they exhibit all the properties prescribed by the ETH for pairs of eigenstates that mix quasimomentum sectors and pairs of eigenstates that do not. Also, the scaled variances $|f_O(\bar E, \omega)|^2$ exhibit the expected diffusive scaling in both sets of matrix elements as $\omega\rightarrow0$. We do find that there are quantitative differences between matrix elements that mix or do not mix quasimomentum sectors; specifically, the scaled variances were found to be generally different, and finite-size effects appear to be stronger in the ones that mix quasimomentum sectors.
 
A much richer behavior was found in interacting integrable models. While the main findings of Ref.~\cite{PRE2019} for translationally invariant observables still apply to observables that break translational symmetry, namely that the off-diagonal matrix elements exhibit skewed log-normal-like distributions and the scaled variances $V_O(\bar E,\omega)$ are well-defined smooth functions, new behaviors were found for symmetry-breaking operators at low frequencies. Most notably, for the operators that have a translationally invariant counterpart that does not break integrability if added as a perturbation to the Hamiltonian, $V_O(\bar E,\omega)$ vanishes as $\omega\rightarrow0$ for matrix elements that do not mix quasimomentum sectors while it approaches a nonvanishing value proportional to $L$ for matrix elements that do. For other observables, $V_O(\bar E,\omega)$ approaches a nonvanishing value proportional to $L$ as $\omega\rightarrow0$ regardless of whether or not the matrix elements mix quasimomentum sectors. However, the low-frequency behavior of $V_O(\bar E,\omega)$ for those other observables is still clearly different between the two sets of matrix elements. For matrix elements that do not mix quasimomentum sectors, there is a dip at low frequencies in $V_O(\bar E,\omega)$ that is absent for those that do. The scaled variances in the latter exhibit a behavior that is qualitatively similar to the one seen in quantum chaotic systems. 

We also showed that, for the observables studied in the integrable XXZ chain (which do not include the spin current~\cite{brenes2020eigenstate}), the lowest frequency scaling of $V_O(\bar E,\omega)$ is consistent with being diffusive regardless of whether the chain is in the easy-plane or easy axis regimes. For integrability-breaking observables, our results suggest that diffusion puts the ultimate limit on the equilibration time in the XXZ chain. In addition, we found a robust frequency regime in which the scaling of $V_O(\bar E,\omega)$ is ballistic for all observables. These results are complementary to the rich recent literature on the interplay between ballistic, superdiffusive, and diffusive spin transport in the XXZ chain and other lattice models~\cite{denardis_2018, Gopalakrishnan2018, Ilievski2018, Gopalakrishnan2019, Gopalakrishnan16250, Bulchandani2020, Agrawal2020, Fava2020} (see Ref.~\cite{2020arXiv200303334B} for a recent review on this topic). 

\acknowledgements
We acknowledge motivating discussions with M. Brenes, A.~Dymarsky, J.~Goold, S.~Gopalakrishnan, R.~Vasseur, and L. Vidmar. This work was supported by the National Science Foundation under Grant No.~PHY-2012145. The computations were carried out at the Institute for Computational and Data Sciences at Penn State.

\appendix

\section{Skewed log-normal-like distributions \\ in the XXZ Chain}\label{appendixskew}

In order to probe whether the skewness observed in Fig.~\ref{fig:8} is a finite-size effect or remains in the thermodynamic limit, we consider $P(\ln|O_{\alpha\beta}|)$ to be a more general function of $\ln|O_{\alpha\beta}|$ than just a Gaussian. Specifically, we take
\begin{equation}\label{eq:ref}
P(\ln|O_{\alpha\beta}|)\propto\exp\left[(\ln D) f\left(\frac{\ln|O_{\alpha\beta}|}{\ln D}\right)\right],
\end{equation}
where $f(x)$ is an unknown concave function (quadratic for the log-normal distribution). This form is motivated by studies of multiplicative noise and multifractals in which similar skewed log-normal-like distributions appear~\cite{referee}. We focus on matrix elements that connect pairs of eigenstates from the same quasimomentum sectors as those are the ones that have been found to exhibit smaller finite-size effects.

\begin{figure}[!t]
\centering \includegraphics[width=\columnwidth]{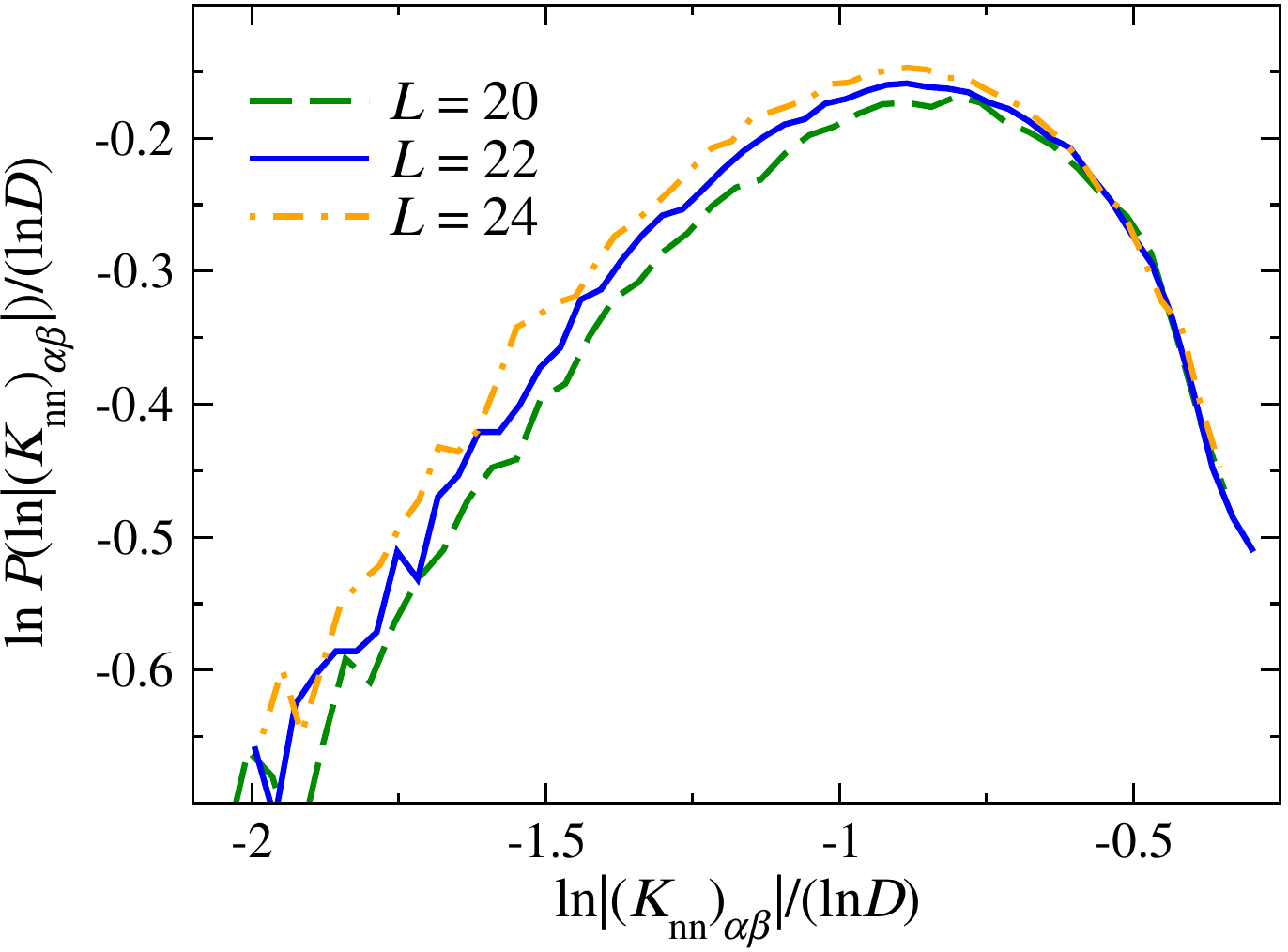}
\vspace{-0.6cm}
\caption{\label{fig:ref} Probability distributions $P(\ln|(K_\text{nn})_{\alpha\beta}|)/(\ln D)$ plotted as functions of $\ln|(K_\text{nn})_{\alpha\beta}| / (\ln D)$ in XXZ chains with $\Delta=0.55$ and $L=20$, 22, and 24. We show results for matrix elements with $k_{\alpha}=k_{\beta}$ selected as explained in Fig.~\ref{fig:8}.}
\end{figure}

In Fig.~\ref{fig:ref} we plot $\ln P(\ln|(K_\text{nn})_{\alpha\beta}|)/(\ln D)$ as a function of $\ln|(K_\text{nn})_{\alpha\beta}|/(\ln D)$ for the three largest chain sizes considered in this work. The data collapse observed suggests that  $P(\ln|(K_\text{nn})_{\alpha\beta}|)$ is described by the ansatz~\eqref{eq:ref} with an $f(x)$ function that is not quadratic, namely, that $P(\ln|(K_\text{nn})_{\alpha\beta}|)$ is a skewed log-normal-like distribution in the thermodynamic limit. We defer finding the $f(x)$ function to future studies. Similar results were obtained for the other integrability-breaking observable $\hat{U}_\text{nn}$ which, like $\hat{K}_\text{nn}$, has a well defined plateau at low frequency in which the scaled variance $V_O(\bar{E},\omega)$ is nonvanishing. 

\section{Ballistic versus Diffusive Scalings\\ in the XXZ Chain}\label{appendix}

\begin{figure}[!t]
\centering \includegraphics[width=\columnwidth]{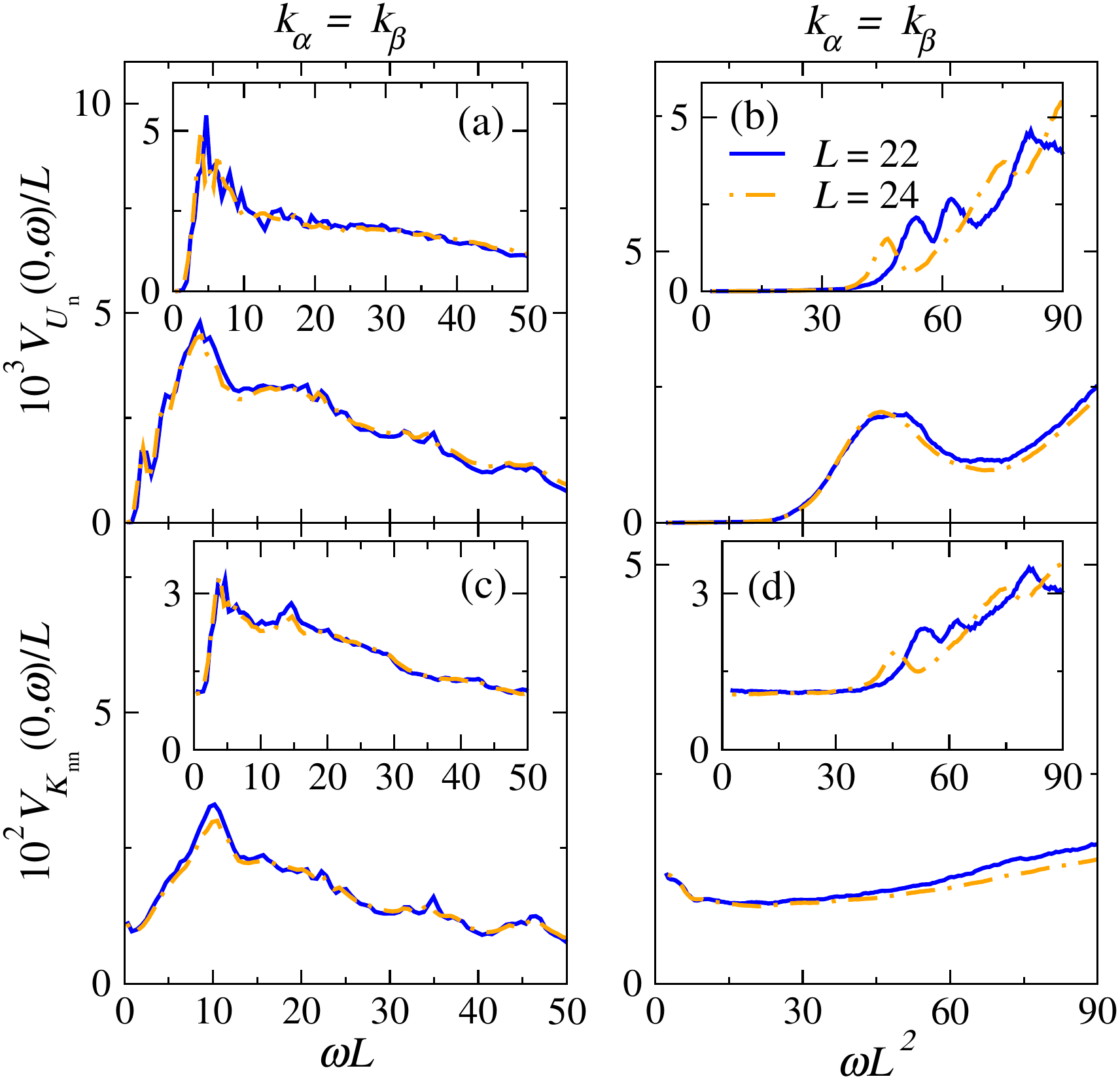}
\vspace{-0.4cm}
\caption{\label{fig:app} Low-frequency plots of the scaled variances $V_O(0,\omega)/L$ vs $\omega L$ [(a), (c)] and vs $\omega L^2$ [(b), (d)] for observables $\hat{U}_\text{n}$ [(a), (b)] and $\hat{K}_\text{nn}$ [(c), (d)] in the integrable XXZ chain with $\Delta=0.55$ (main panels) and 1.1 (insets). The results are for the two largest chain sizes studied ($L=22$ and 24) for matrix elements between pairs of energy eigenstates with $k_{\alpha}=k_{\beta}$. All pairs of eigenstates satisfy $|\bar{E}|/L\leq 0.025$. The averages $\overline{|O_{\alpha\beta}|^2}$ in (a) and (c) were coarse-grained in windows of width $\delta \omega = 0.025$. The running averages $\overline{|O_{\alpha\beta}|^2}$ in (b) and (d) were calculated in windows of width $\delta \omega=0.009$ centered at points separated by $\Delta \omega=0.001$.}
\end{figure}

Here we show the low-frequency behavior of the scaled variances $V_{U_\text{n}}(0,\omega)/L$ [Figs.~\ref{fig:app}(a) and~\ref{fig:app}(b)] and $V_{K_\text{nn}}(0,\omega)/L$ [Fig.~\ref{fig:app}(c) and~\ref{fig:app}(d)] plotted versus $\omega L$ [Figs.~\ref{fig:app}(a) and~\ref{fig:app}(c)] and versus $\omega L^2$ [Figs.~\ref{fig:app}(b) and~\ref{fig:app}(d)] side-by-side for the two largest (integrable) XXZ chains studied ($L=22$ and $L=24$). The main panels show results for $\Delta=0.55$ while the insets show results for $\Delta=1.1$. Figure~\ref{fig:app} makes apparent that the data collapses discussed in the main text improve with increasing chain size. Also, plotting only two chain sizes in Fig.~\ref{fig:app} allows one to better see the level of detail at which the data collapses occur, including the various features in the scaled variances whose location remains system-size independent in the plots versus $\omega L$ and versus $\omega L^2$.

\newpage

\bibliographystyle{biblev1}
\bibliography{refs}

\end{document}